\documentclass[prb,footnoteinbib,showkeys,showpacs,twocolumn]{revtex4}
\usepackage{amsmath,amssymb,epic,eepic}
\usepackage{graphicx}
\usepackage{hyperref}

\begin{document}

\title{Fractionalization of minimal excitations in integer quantum Hall edge channels}

\author{Ch. Grenier}

\address{Centre de Physique Th\'eorique (CPHT), Ecole Polytechnique, 91128 Palaiseau Cedex - France}

\author{J. Dubois}
\author{T. Jullien} 
\author{P. Roulleau}
\author{D.C. Glattli}

\affiliation{Nanoelectronic Group, Service de Physique de l'Etat Condens\'e, CEA Saclay, F-91191 Gif-Sur-Yvette, France}

\author{P. Degiovanni}

\affiliation{Universit\'e de Lyon, F\'ed\'eration de Physique Andr\'e Marie Amp\`ere,\\
CNRS - Laboratoire de Physique de l'Ecole Normale Sup\'erieure de Lyon,\\
46 All\'ee d'Italie, 69364 Lyon Cedex 07, France}

\begin{abstract}
A theoretical study of the single electron coherence properties of Lorentzian and rectangular
pulses is presented. By combining bosonization and the Floquet scattering
approach, the effect of interactions on a periodic source of voltage pulses is computed exactly.
When such excitations are injected into one of the channels of a system of two copropagating quantum Hall edge channels, 
they fractionalize into pulses whose charge and shape reflects the properties of interactions. We show that the 
dependence of fractionalization induced electron/hole pair production in the pulses amplitude contains 
clear signatures of the fractionalization of the individual excitations. We propose an experimental
setup combining a source of Lorentzian pulses and an Hanbury Brown and Twiss interferometer to measure
interaction induced electron/hole pair production and more generally to reconstruct single electron coherence
of these excitations before and after their fractionalization. 
\end{abstract}



\maketitle

\section{Introduction}

Coherence of electrons propagating along the edge channels of a 2DEG has been demonstrated a few years ago in electronic
Mach-Zehnder interferometry experiment\cite{Ji:2003-1}. This breakthrough has triggered a remarkable thread of 
experimental\cite{Neder:2007-4,Roulleau:2007-2,Roulleau:2008-1,Roulleau:2008-2,Litvin:2007-1} and theoretical 
studies\cite{Chalker:2007-1,Levkivskyi:2008-1,Kovrizhin:2009-1,Neuenhahn:2008-1,Neder:2007-3}.
Recently, the successive partitioning of single electron and single hole excitations emitted by a mesoscopic capacitor\cite{Feve:2007-1,Mahe:2010-1}
using an electronic Hanbury Brown and Twiss setup\cite{Henny:1999-1,Oliver:1999-1} 
has been demonstrated\cite{Bocquillon:2012-1}.
This opens the way to quantitative studies of decoherence and relaxation of single electron excitations in electronic 
systems\cite{Degio:2011-1}.  

For example, on demand single electron source\cite{Feve:2007-1,Leicht:2011-1,Hohls:2012-1} in quantum Hall edge channels
or in other non-chiral systems\cite{Talyanskii:1997-1,Blumenthal:2007-1,Ahlers:2006-1,Kaestner:2008-1,Fujiwara:2008-1}
could be used to study the problem of quasi particle 
relaxation originally considered by  Landau in the Fermi liquid theory\cite{Landau:1959-1,Nozieres-Pines} and 
recently solved non perturbatively in integer quantum Hall
edge channels\cite{Degio:2009-1}. However, considering other types
of coherent single electron excitations less sensitive to decoherence and simpler to generate is certainly of great
interest for the field. 

Levitov, Ivanov,
Lee and Lesovik\cite{Levitov:1996-1,Ivanov:1997-1}, have proposed a way to generate clean current
pulses carrying single to few coherent electron excitations. Elaborating over these pioneering works, Keeling, Klich
and Levitov\cite{Keeling:2006-1} have extended this to the case of fractional excitations in FQHE and Tomonaga-Luttinger liquids.
Contrary to the excitations generated by the mesoscopic capacitor \cite{Feve:2007-1}, 
these few electron pulses are time resolved instead of being energy resolved. Moreover, pulses that carry more
than an elementary charge inject a coherent wave packet of $n$ electrons. This opens the way to entangle several
electrons and to probe the full counting statistics\cite{Levitov:1996-1} with a finite number of particles\cite{Hassler:2008-1}.

\medskip

In this paper, we show that these  electron pulses are very convenient probes of fractionalization\cite{Su:1979-1} induced by electron/electron interactions in
quantum Hall edge channels. To reach this conclusion, we have studied their decoherence and relaxation due to electron/electron
interactions in a system of copropagating edge channels at filling fraction $\nu=2$ recently used to study 
energy relaxation\cite{Altimiras:2010-1,LeSueur:2010-1,Altimiras:2010-2}. 
In this system, neutral/charged mode separation\cite{Bocquillon:2012-2} leads to fractionalization of a Lorentzian pulse propagating within one 
of the edge channel into two components. As in the case of non chiral 1D systems\cite{Safi:1995-1,Pham:2000-1,Trauzettel:2004-1,Steinberg:2008-1}, 
each of the two resulting pulses carries a fraction of the charge directly related to the interaction strength. 

In a recent work\cite{Neder:2012-1}, Neder has proposed to detect a signature of fractionalization
of a continuous flow of electrons in the shot noise which is known to count the number
of all excitations (electrons and holes) when the pulses are partitioned
at an electronic beam splitter\cite{Levitov:1996-1,Degio:2010-4,Dubois:2012-1}. In this paper, we propose to study the phenomenon 
of fractionalization at the single excitation level using Lorentzian pulses and we discuss the conditions of its observability as well
as the information that could be extracted from noise measurements.

The experimental setup we propose combines a Lorentzian pulse source with an electronic Hanbury Brown and 
Twiss (HBT) interferometer. With our design, current noise and high frequency admittance measurements
could then be combined into quantitative tests of the description of fractionalization in terms
of edge magnetoplasmon scattering, recently directly measured in a $\nu=2$ edge channel system\cite{Bocquillon:2012-2}, 
and which plays a role in electronic relaxation in quantum Hall 
edge channels\cite{Degio:2009-1,Degio:2010-1,Levkivskyi:2012-1} as well as decoherence in 
MZI interferometers\cite{Levkivskyi:2008-1,Neuenhahn:2009-1,Kovrizhin:2009-1}. 

In this perspective, we present predictions for the interaction induced electron/hole pair production 
for Lorentzian and rectangular pulses.  In the case of (screened) short range interactions,
we  predict stroboscopic decoherence and revivals of Lorentzian and rectangular pulses as a function of copropagation length. We identify clear
signatures of fractionalization in the behavior of electron/hole production as a function of the amplitude of the pulses and of the copropagation
distance. Taking into account finite temperature, we show that these signatures could be observed in realistic experiments. 
However, in the case of long range interactions, we show that the signature of fractionalization disappears due to the dispersion
of the edge magnetoplasmon eigenmodes. 

\medskip

This paper is structured as follows: in section \ref{sec:source}, we briefly discuss the single electron coherence generated by a periodic
source of voltage pulses. Then,  in section \ref{sec:decoherence}, we show how to describe
the effect of electron/electron interactions on these pulses by combining bosonization with the Floquet scattering theory. This leads
us to quantitative predictions for interaction induced electron/hole pair production in section \ref{sec:HBT}. 

\section{Generating and characterizing single electron pulses}
\label{sec:source}

\subsection{Voltage pulse generated states}
\label{sec:source:single-shot}

Pure electronic excitations
can be created into a chiral edge channel without exciting any electron/hole pair by applying suitable voltage pulses to an Ohmic 
contact\cite{Levitov:1996-1,Keeling:2006-1}.  
Since the average current $\langle i(t)\rangle$ injected into a single chiral edge channel 
by a time dependent drive $V(t)$ applied to an Ohmic contact is $\langle i(t)\rangle=(e^2/h)V(t)$, 
the voltage drive must satisfy the quantization condition such that $e\int V(t)\,dt$ be a multiple of $h$: the average emitted charge 
$\int \langle i(t)\rangle\,dt$ is then a multiple of the elementary charge $-e$.

\medskip

But this quantization condition is not sufficient: an important result by Levitov, Lee and Lesovik\cite{Levitov:1996-1} states that in
order to generate a state involving pure electronic excitations, the voltage drive $V(t)$ must be a sum of elementary Lorentzian pulses 
carrying exactly one electron. 

\subsubsection{The integer charge Lorentzian pulse as a Slater determinant}
\label{sec:source:fermionization}

Let us start by considering a single Lorentzian voltage pulse centered at time $t=0$ such that $-e\int V(t)dt=\alpha h$ where in this paragraph, $\alpha$ is a positive integer $n$:
\begin{equation}
\label{eq:coherence:Levitov:single-electron-pulse}
V(t)=-\frac{2\alpha\hbar}{e}\,\frac{\tau_0}{t^2+\tau_{0}^2}
\end{equation}
where $\tau_{0}$ is the duration of the pulse. 

\medskip

The appropriate way to characterize 
single electron coherence in a many body system is to consider the two particle Green's function at a given 
position\cite{Degio:2011-1,Degio:2010-4}:
\begin{equation}
\mathcal{G}^{(e)}(t,t')=
\langle \psi^\dagger(x,t')\psi(x,t)\rangle
\end{equation}
where the bracket denotes a quantum average in the presence of the electronic sources.
This function plays the same role as the first order coherence in quantum optics\cite{Glauber:1963-1}. 
This single electron coherence characterizes the electronic quantum coherence at the single particle level: in the absence of
interactions, the outcoming current from an electronic Mach-Zenhder interferometer (MZI) can indeed be expressed in terms of the incoming
single electron coherence\cite{Haack:2011-1,Haack:2012-2}. Exactly as in quantum optics\cite{Lvovsky:2009-1}, 
Hanbury Brown and Twiss interferometry provides a way to reconstruct
the single electron coherence from current noise measurements\cite{Degio:2010-4}  or, at least, to compare excitations of quantum states
emitted by two different sources\cite{Moskalets:2010-1} via an Hong Ou Mandel type experiment.

\medskip

Being submitted
to this voltage drive within the Ohmic contact, each electron accumulates an electric phase which then appears
in the temporal coherence properties of electrons emitted by the Ohmic contact. In the presence of a
general time dependent external drive $V(t)$, the single electron coherence of electrons $\mathcal{G}^{(e)}(t,t')$ is given by
\begin{equation}
\label{eq:coherence:driven}
\mathcal{G}^{(e)}(t,t')=\exp{\left(\frac{ie}{\hbar}\int_{t'}^{t}V(\tau)\,d\tau\right)}\,\mathcal{G}^{(e)}_{\mu}(t,t')\,.
\end{equation}
where $\mathcal{G}^{(e)}_{\mu}$ denotes the single electron coherence
for the edge channel at chemical potential $\mu$. 
For the Lorentzian pulse carrying $n$ electrons, eq.~\eqref{eq:coherence:driven} becomes:
\begin{equation}
\label{eq:coherence:Levitov:single-electron-coherence}
\mathcal{G}^{(e)}(t,t')=\left(\frac{t+i\tau_0}{t-i\tau_0}\,\frac{t'-i\tau_0}{t'+i\tau_0}\right)^{n}\,
\mathcal{G}_{\mu}^{(e)}(t,t')\,.
\end{equation}
At zero temperature, the contribution $\Delta_\mu\mathcal{G}^{(e)}=\mathcal{G}^{(e)}-\mathcal{G}^{(e)}_{\mu}$ of the pulse to single electron
coherence, expressed in the frequency domain, is given by:
\begin{eqnarray}
\label{eq:coherence:Levitov:Laguerre}
v_F\,\Delta_\mu\mathcal{G}^{(e)}(\omega_{+},\omega_{-}) & = & 4\pi\tau_0\,
\Theta(\omega_+)\Theta(\omega_-)\,e^{-(\omega_{+}+\omega_{-})\tau_0}\nonumber \\
& \times & \sum_{k=0}^{n-1}L_{k}(2\omega_{+}\tau_{0})L_{k}(2\omega_{-}\tau_{0})
\end{eqnarray}
where $\omega_{\pm}$ are respectively conjugate to $t$ and $t'$ and $L_{k}$ denotes the $k$th Laguerre polynomial. The vanishing of 
$\Delta_\mu\mathcal{G}^{(e)}(\omega_{+},\omega_{-})$ when either $\omega_{+}$ or $\omega_{-}$ is negative signals the absence of hole
excitations in this $n$-electron Lorentzian pulse. 

In fact, expression \eqref{eq:coherence:Levitov:Laguerre} is nothing but the single electron coherence
for a many body state obtained by $n$ electrons in a Slater determinant built from an orthonormal family or $n$ positive energy states
on top of the Fermi sea $|F_\mu\rangle$ at chemical potential $\mu$. More precisely, 
denoting by $(\varphi_k)_k$ these $n$ wave functions, Wick's theorem directly implies that the single
electron coherence of the many body state 
\begin{equation}
|\Psi_\mu[(\varphi_k)_k]\rangle=\prod_{k=1}^n\psi^\dagger[\varphi_k] \,|F_\mu\rangle
\end{equation}
obtained by adding $n$ electrons in the $n$ single particle states
$\varphi_1,\ldots,\varphi_n$ is given in the space domain by
\begin{equation}
\mathcal{G}^{(e)}(x,y)=\mathcal{G}^{(e)}_\mu(x,y)+\sum_{k=1}^n\varphi_k(x)\varphi_k(y)^*\,.
\end{equation}
Since, for a many body state generated from $|F_\mu\rangle$ by the application of an external
time dependent voltage drive, single electron coherence determines all the electronic correlation functions, this
shows that the many body state generated by the 
$n$-electron pulse \eqref{eq:coherence:Levitov:single-electron-pulse} 
is obtained  by adding on top of the Fermi sea the $n$ normed and mutually orthogonal wavepackets which are ($0\leq k\leq n-1$):
\begin{equation}
\label{eq:coherence:Levitov:wavefunctions}
\varphi_{k}^{(\tau_0)}(\omega)=\sqrt{2\tau_0}\,\Theta(\omega)e^{-\omega\tau_{0}}L_{k-1}(2\omega\tau_0)\,.
\end{equation}
Consequently, the $n$-particle wavefunction describing the added electrons is a Slater determinant built
from the wavefunctions $\varphi_k^{(\tau_{0})}$ ($1\leq k\leq n$). Let us mention
that this result can also be reached by considering the algebra of operators
associated with the creation of a single $\alpha=1$ Lorentzian pulse along the lines followed by 
Keeling, Klich and Levitov\cite{Keeling:2006-1}. 

\subsubsection{Floquet approach to a periodic trains of pulses}
\label{sec:source:periodic}

Because single shot detection of single electron excitations is not available today, we have to resort on statistical measurements
to characterize these excitations. Therefore, in practice, we shall consider a periodic source 
delivering an infinite train of periodically spaced pulses:
\begin{equation}
\label{eq:periodic-voltage}
V(t) =  \sum_{m=-\infty}^{+\infty}V_{\mathrm{p}}(t-mT)
\end{equation}
where $V_{\mathrm{p}}(t)$ denotes a single pulse.
As discussed in \cite{Dubois:2012-1}, a periodic source can be used to generate a train of 
pulses carrying a non integer charge since the amplitude of the voltage drive can be varied continuously. In this case, each 
pulse can no longer be viewed as obtained from the Fermi sea by adding only electron or hole excitations. On the contrary, it should be understood as a collective
excitation of the electronic fluid that contains both electron and hole excitations. 

\medskip

Since the voltage drive \eqref{eq:periodic-voltage} is $T$-periodic, the single particle coherence \eqref{eq:coherence:driven} is also $T$-periodic. 
Considering $V_{\mathrm{ac}}(t)=V(t)-\alpha hf/e$ the AC part of the voltage drive, the associated phase accumulated in the time interval $[0,t]$ is periodic 
and can thus be decomposed in a Fourier series:
\begin{equation}
\label{eq:Floquet:phase}
\exp{\left(\frac{ie}{\hbar}\int_{0}^{t}V_{\mathrm{ac}}(\tau)\,d\tau\right)}=\sum_{l=-\infty}^{+\infty}C_{l}[V_{\mathrm{ac}}]\,e^{-2\pi iflt}\,.
\end{equation}
As discussed in our previous work\cite{Dubois:2012-1}, 
the Fourier coefficients $C_l[V_{\mathrm{ac}}]$ are the Floquet scattering amplitudes\cite{Moskalets:2002-1,Moskalets:book} for electrons: for 
$l>0$, $C_l[V_{\mathrm{ac}}]$ is the amplitude for a free electron to absorb $n$ photons of energy $hf$
whereas for $l<0$ it is the amplitude for a free electron to emit $n$ photons. Finally, $C_0[V_{\mathrm{ac}}]$ is the amplitude for the electron
not to absorb nor emit any photon. Note that the probabilities for photo assisted transitions $P_l[V_{\mathrm{ac}}]=|C_l[V_{\mathrm{ac}}]|^2$ 
sum to unity: $\sum_{l}P_l[V_{\mathrm{ac}}]=1$. 

Using these Floquet amplitudes, the single electron coherence can then be decomposed as\cite{Degio:2010-4}:
\begin{equation}
\label{eq:coherence:Fourier}
\mathcal{G}^{(e)}(t,t')=\sum_{n=-\infty}^{+\infty}e^{-2\pi i n f \bar{t}}\int \mathcal{G}_{n}^{(e)}(\omega)\,e^{-i\omega(t-t')}\frac{d\omega}{2\pi}\,.
\end{equation}
Equations \eqref{eq:coherence:driven} and \eqref{eq:Floquet:phase} then lead to\cite{Dubois:2012-1}:
\begin{equation}
\label{eq:Floquet:coherence}
v_{F}\mathcal{G}_{n}^{(e)}(\omega)=
\sum_{l=-\infty}^{+\infty}C_{n+l}[V_{\mathrm{ac}}]\,C_{l}[V_{\mathrm{ac}}]^*\,f_{\bar{\mu},T_{\mathrm{el}}}(\omega-(n+2l)\pi f)\,.
\end{equation}
where $f_{\bar{\mu},T_{\mathrm{el}}}$ denotes the Fermi distribution at chemical potential $\bar{\mu}=\mu+\alpha hf$ and electronic
temperature $T_{\mathrm{el}}$. This shift arises from the dc component of the voltage $V(t)$. 
Eq.~\eqref{eq:Floquet:coherence} expresses that the single electron coherence is built by summing the photo-generated single 
particle coherence over all independent electrons belonging to the shifted Fermi sea. 

\medskip

The nature of single particle excitations contained within the pulse train can be read from the $\mathcal{G}^{(e)}$ in the frequency domain
where it is a function of the two angular frequencies $\omega_+$ and $\omega_-$ respectively conjugated to $t$ and $t'$. 
For a periodic source, $\mathcal{G}^{(e)}_n(\omega)$ defined by \eqref{eq:coherence:Fourier} describes the single electron coherence 
for $\omega_\pm = \omega\pm \pi nf$: the contribution of electronic excitations (with respect to $\mu=0)$
comes from $\omega \geq \pi f|n|$ whereas the contribution from hole excitations comes from $\omega \leq -\pi f|n|$. 
Contributions to $\mathcal{G}^{(e)}$ in the two quadrants $|\omega|< \pi f|n|$
corresponding to $\omega_+\omega_-<0$ stem from electron hole coherences\cite{Degio:2010-4} and can be shown to be
responsible for  a positive peak in current correlations in an Hong-Ou-Mandel collision experiment\cite{Jonckheere:2012-1}.

In the frequency domain, the diagonal part $\mathcal{G}^{(e)}_0(\omega)$ of the single electron coherence
is nothing but the electron distribution function. For a periodic train of voltage pulses, it can be expressed in terms of the photo-assisted 
transition probabilities\cite{Dubois:2012-1}:
\begin{equation}
\label{eq:Floquet:population}
f_e(\omega)=v_F\mathcal{G}_0^{(e)}(\omega)=
\sum_{l=-\infty}^{+\infty} P_l[V_{\mathrm{ac}}]\,f_{\bar{\mu},T_{\mathrm{el}}}(\omega-2\pi lf)
\end{equation}
As we will see in section \ref{sec:decoherence}, eq.~\eqref{eq:Floquet:coherence} and \eqref{eq:Floquet:population} can also be used
to compute this quantity in the presence of electron-electron interactions. But before discussing interaction effects, lets us introduce the
voltage drives considered in the present paper.

\subsubsection{Lorentzian and rectangular pulse trains}

These expressions are valid for any periodic voltage drive and, as in our recent paper\cite{Dubois:2012-1}, three examples will be considered here: 
the sine periodic drive and periodic trains of Lorentzian and rectangular pulses. These voltage drives are always decomposed into a dc part $V_{\mathrm{dc}}$
and an ac component. The amplitude of the ac component 
is directly related to the amplitude of the dc part which determines 
the average charge injected per period $-\alpha e$. 

\medskip

\begin{figure}
\includegraphics[width=8cm]{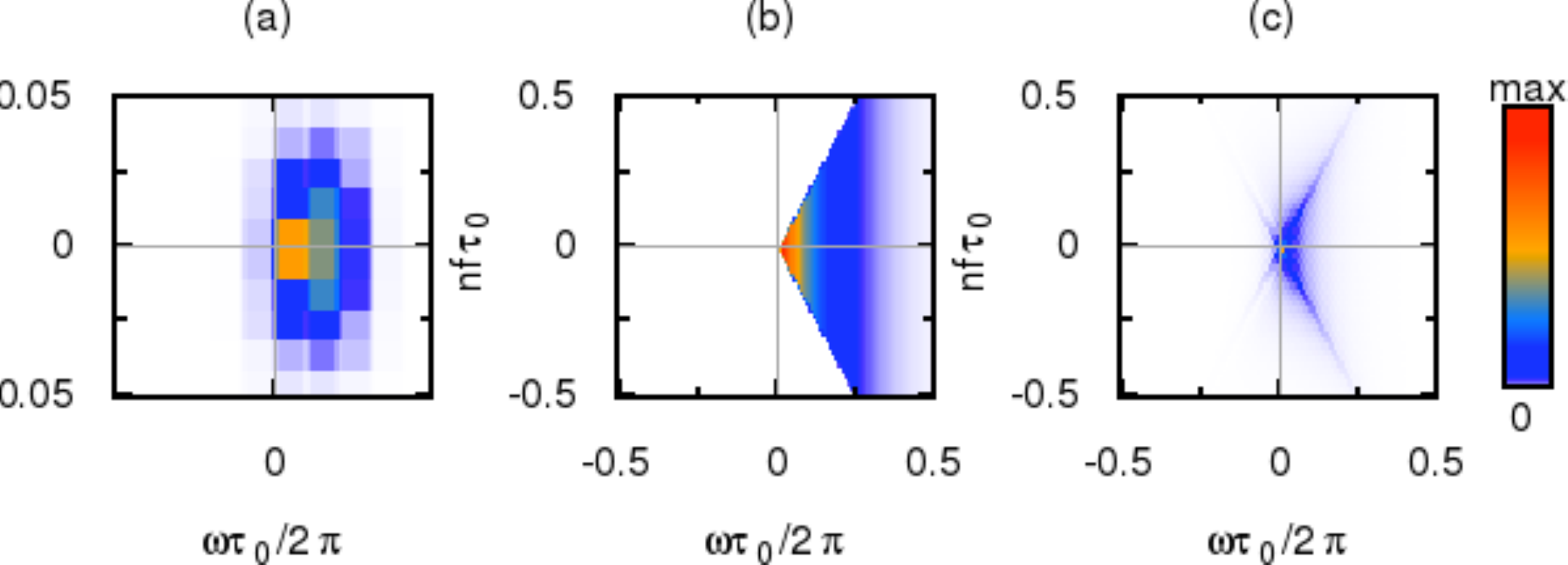}
\caption{\label{fig:coherence:various-signals} (Color online) Modulus of the excess single electron coherence in the frequency domain for various pulses at zero temperature and $\alpha=1$: (a) sinusoidal drive (b) Lorentzian drive $f\tau_0=0.01$ (c) rectangular drive with $f\tau_0=0.01$. Electronic (reps. hole) excitations are located in the $\omega\geq \pi |n|f$ (reps. $\omega\leq -\pi f|n|$) quadrant whereas the off diagonal $|\omega|\leq \pi f|n|$ quadrants correspond to electron/hole coherences.}
\end{figure}

For the sine voltage, $V(t)=V_{\mathrm{dc}}(1-\cos{(2\pi ft)})$ where $V_{\mathrm{dc}}=-\alpha hf/e$. The photo-assisted transition 
amplitudes have the well known expression\cite{Tien:1963-1} $C_n=J_n(\alpha)$. The resulting excess single electron coherence 
$\Delta \mathcal{G}^{(e)}$ in
the frequency domain is depicted on graph (a) of figure \ref{fig:coherence:various-signals}. As expected, it is located close
to the Fermi surface and includes both electron and hole contributions as well as electron/hole coherences.
Note that the photo-emission and photo-absorption probabilities 
are equal since the sine drive is symmetric with respect to its average $V(t+T/2)+V(t)=V_{\mathrm{dc}}$. This is clearly not the case for the two other families
of pulses considered here.

\medskip

For each of these two families, the physics depends on two parameters: the injected charge $\alpha$
in units of $-e$ and $f\tau_0$ which characterizes the compacity of the pulse train, {\it i.e.} the size of each pulse compared to their spacing.
The physics of well individualized pulses is best probed at low compacity\cite{Dubois:2012-1} $f\tau_0\ll 1$ whereas at high
compacity, a shifted Fermi sea is recovered due to the Pauli exclusion principle. To illustrate this regimes, many numerical results 
will be presented for $f\tau_0=0.01$ to shed light on the physics at the level of a single pulse.

In realistic experiments, typical values of $\tau_0$ are of the order of 
40~ps and the drive frequency is typically
$f=5\ \mathrm{GHz}$ thus leading to $f\tau_0\simeq 0.2$. Then a 50~mK electronic temperature corresponds to $k_BT_{\mathrm{el}}/hf\simeq 0.2$.
Reaching lower values of $\tau_0$ and higher frequency drives so that $k_BT_{\mathrm{el}}/hf$ would be more favorable
at fixed $f\tau_0$ would require the use of optical techniques\cite{Kato:2005-1}. Consequently, when discussing the observability of fractionalization,
we will choose $f\tau_0=0.1$, a value within reach of current technologies.

\medskip

As discussed in our recent paper\cite{Dubois:2012-1}, in the low $f\tau_0$ regime and
at zero temperature, both Lorentzian and square pulses exhibit a minimal number of electron/hole pairs
for all integer values of the charge $\alpha$ carried by each pulse. This seems to be the case
for arbitrary waveforms as noticed by several authors\cite{Vanevic:2008-1,Gabelli:2012-1}. But in the case of the Lorenztian pulses, as predicted by
Levitov, Lee and Lesovik\cite{Levitov:1996-1}, the
the number of hole (reps. electron) excitations vanishes for positive (reps. negative) integer 
values of $\alpha$ whereas for all other waveforms, it presents a non zero minimum. 

The difference in these behaviors reflects the presence of hole excitations for non-Lorentzian pulses for positive integer $\alpha$. 
Figure \ref{fig:coherence:various-signals} clearly shows the difference between a Lorentzian pulse with $\alpha=1$ (graph (b)) and a rectangular pulse
with the same $\alpha$ and $f\tau_0$ (graph (c)): the single electron coherence for the Lorentzian pulse is localized
within the electron quadrant $\omega \geq \pi f|n|$ whereas, for rectangular pulses, hole excitations as well as electron-hole
coherences are present.

As can be seen 
from \eqref{eq:Floquet:population}, this can be traced back to the properties of the Floquet amplitudes $C_l[V_{\mathrm{ac}}]$ for $l+\alpha<0$
whose analytical expressions  for a periodic train of Lorentzian and rectangular pulses are given in appendix \ref{appendix:analytics}. 
These amplitudes vanish for Lorentzian pulses for positive integer $\alpha$. As expected, the expression for the single electron coherence 
\eqref{eq:Floquet:coherence} shows that when the Floquet amplitudes $C_l[V_{\mathrm{ac}}]$ for $l+\alpha<0$ vanish, the excess single electron coherence
due to the source has non zero contributions only in the electron quadrant\cite{Degio:2010-4}: this reflects the purely electronic nature of the
wave packets generated by the source in this case. For negative $\alpha$ the discussion goes along the same line with holes replacing the electrons.

\subsection{Counting electron and hole pairs}
\label{sec:characterization}

\subsubsection{Proposed experimental setup}
\label{sec:characterization:setup}

\begin{figure}
\includegraphics[width=8cm]{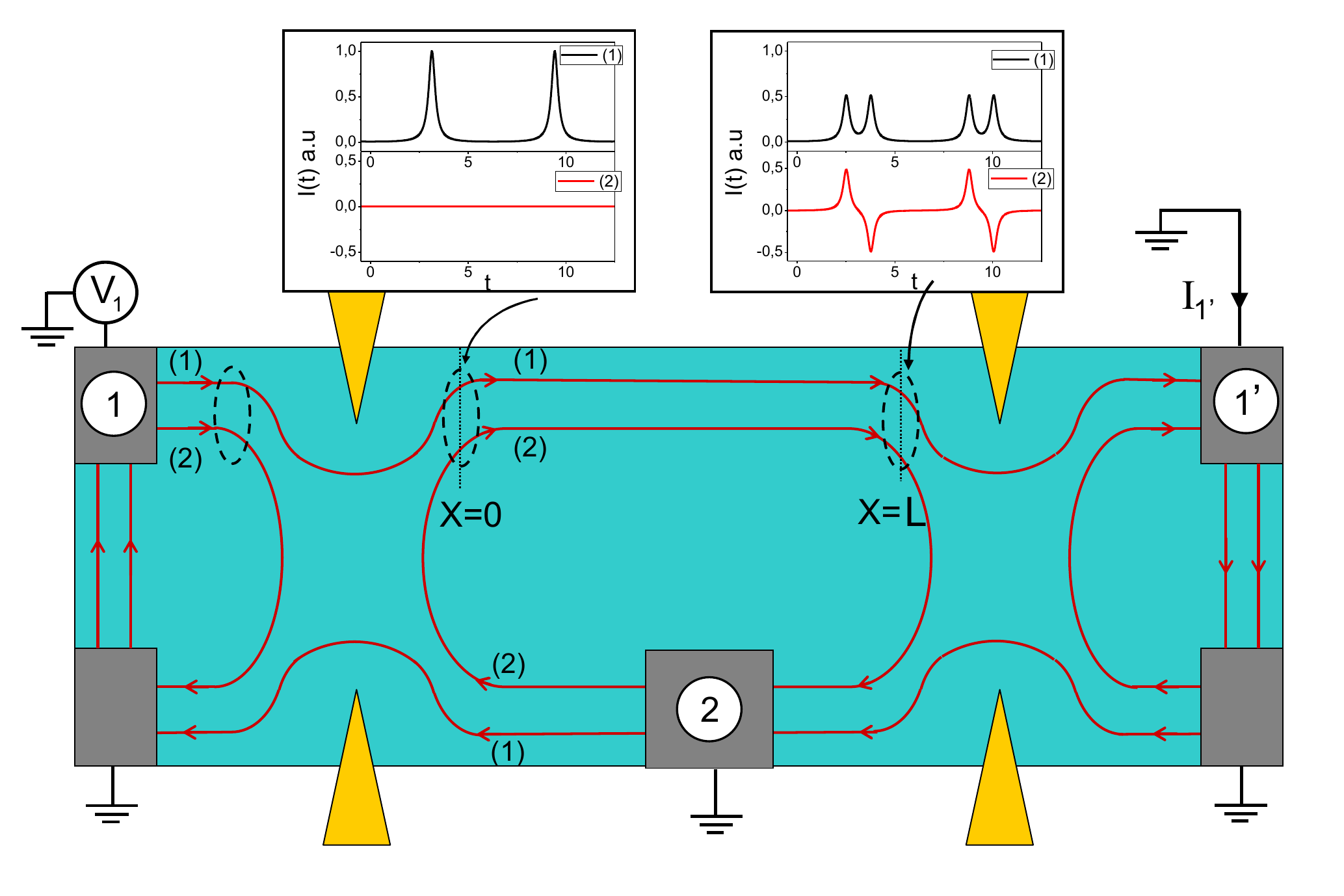}
\caption{\label{fig:HBT} (Color online) Hanbury-Brown Twiss setup for the $\nu=2$ system: edge channels $(1)$ and $(2)$ interact between $x=0$ and $x=l$. 
The Ohmic contact $1$ is driven by the $T$-periodic voltage 
\eqref{eq:periodic-voltage} whereas the Ohmic contact $2$ is at a fixed chemical potential. 
One then measures the average and the low frequency noise of the current  $I_{1'}$ at contact $1'$. 
Depending on the polarizations of the two QPCs this setup
performs an HBT experiment right after the source $(1')$ or after excitations within the outer channel have co-propagated
along the inner edge channel over a distance $l$.}
\end{figure}

Let us now describe the experimental setup that we propose to study interaction induced electron/hole pair production using shot
noise measurements in an Hanbury Brown and Twiss interferometer\cite{Degio:2010-4,Bocquillon:2012-1,Dubois:2012-1}.
The setup depicted on figure \ref{fig:HBT} consists of two quantum point contacts (QPC) located at both ends of
a region where two chiral edge channels interact during their copropagation. Excitations are generated by driving the
Ohmic contact (1) and noise and current measurement at leads (1') and (2') can be performed.
In a previous work, we have proposed this setup to study interchannel
frequency dependent energy transfer using finite frequency noise measurements\cite{Degio:2010-1}. But as we shall now
explain, playing with the polarizations of both QPCs enables us to compare electron/hole production for different propagation distances
and thus to study the effects of interactions on this quantity. This idea has also been recently used to measure edge magnetoplasmon scattering
in a $\nu=2$ edge channel system\cite{Bocquillon:2012-2}.

\medskip

To begin with, let us assume that the first QPC is polarized so that the inner edge channel is totally reflected and the outer one partially transmitted whereas 
the second QPC is unpolarized so that both channels are transmitted. This performs an HBT experiment at the first QPC in order 
to characterize electronic excitations propagating in the outer edge channel right for a small copropagation distance $l$.

Next, the first quantum point contact can also be polarized to inject a periodic train of single electron pulses into 
the outer edge channel and an equilibrium state into the inner edge channel. When the second QPC is  polarized to totally 
reflect the inner edge channel and to partition the outer one, an HBT experiment is performed on the outer edge channel
after copropagation along the inner edge channel over a large distance. 

\medskip

Note that in the setup proposed here, there are two easily accessible experimental controls. The first one is the amplitude of the pulse which determines 
the injected charge. The second one is the driving frequency of the source.
In principle, one could also design a sample that allows copropagation over different distances $l$ as in the recent experiment by F. Pierre 
and his collaborators\cite{LeSueur:2010-1,Altimiras:2010-2}. Combining the use of an appropriate geometry with the variation of the drive
frequency keeping the compacity $f\tau_0$ of the source constant within the limitations of the pulse generator, 
one could thus explore the $(\alpha,l)$ plane and study how electron/hole pair production depends on these two parameters.

But before presenting our predictions concerning the effect of interactions, we shall now review how low frequency current noise 
 in the HBT geometry is related to electron/hole pair production.

\subsubsection{Electron/hole pair production}
\label{sec:characterization:hbt}

In an HBT experiment, the current noise in one of the outcoming channels of a quantum point contact 
is measured at low frequency:
\begin{equation}
S_{1',1'}^{\mathrm{exp}}=\int 
\overline{\langle I_{1'}(\bar{t}+\tau/2)\,I_{1'}(\bar{t}-\tau/2)\rangle_c}^{\bar{t}}\,d\tau
\end{equation}
where an average over $\bar{t}=(t+t')/2$ has been taken. The outcoming current noise $S_{1',1'}^{\mathrm{exp}}$ 
contains contributions of the incoming current noise from the two incoming channel and a contribution associated with
two particle interference effects. The latter can be expressed in terms of the single electron coherences
of the two incoming channels\cite{Degio:2010-4}. At zero temperature, when the AC drive is switched off, $S_{1',1'}^{\mathrm{exp}}$ is still
non zero because of the partition noise of electrons due to the DC bias $V_{\mathrm{dc}}=-\alpha hf/e$. It is thus convenient to
consider the excess noise $\Delta S_{1',1'}^{\mathrm{exp}}=
(S_{1',1'}^{\mathrm{exp}})_{\mathrm{ac}+\mathrm{dc}}-(S_{1',1'}^{\mathrm{exp}})_{\mathrm{dc}}$ 
due to the AC part of the voltage drive: this excess noise solely arises
from the partitioning of electron and hole excitations associated with the AC voltage drive. As discussed in the framework of 
photo-assisted noise\cite{Lesovik:1994-1,Pedersen:1998-2,Rychkov:2005-1,Schoelkopf:1998-2,Reydellet:2003-1,Dubois:2012-1} 
and as we shall now briefly recall in the electron quantum optics language, this excess noise counts the electron/hole pair production.
This procedure has originally been proposed by M. Vanevic, Y. V. Nazarov, and W. Belzig\cite{Vanevic:2008-1}
and discussed extensively in our previous work\cite{Dubois:2012-1}. 

\medskip

Since an Ohmic contact driven by an AC voltage has the same current fluctuations as in the absence of AC drive (even at non zero 
temperature), the excess current noise
$\Delta S_{1',1'}^{\mathrm{exp}}$ due to the AC component of the drive voltage is given by the excess contribution arising from two particle
interferences. 
Assuming that the QPC behaves as  an energy independent electronic beam splitter with reflexion and transmission probabilities $\mathcal{R}$ and $\mathcal{T}$, 
$\Delta S_{1',1'}^{\mathrm{exp}}/\mathcal{R}\mathcal{T}$ is the overlap of the single electron and single hole excess coherences of the 
source due to the AC drive $\Delta\mathcal{G}_{1}^{(e/h)}(t,t')
=\mathcal{G}_1^{(e/h)}(t,t')-\mathcal{G}^{(e/h)}_{\bar{\mu},T_{\mathrm{el}}}(t,t')$ with the electron and hole coherences $\mathcal{G}^{(e/h)}_{3'}(t,t')$ 
emitted into the other incoming channel\cite{Degio:2010-4}:
\begin{equation}
\label{eq:HBT:overlap:delta} 
\frac{\Delta S_{1',1'}^{\mathrm{exp}}}{\mathcal{R}\mathcal{T}}=(ev_F)^2\int \overline{(\Delta\mathcal{G}^{(e)}_{1}\mathcal{G}^{(h)}_{3'}+
\Delta\mathcal{G}^{(h)}_{1}\mathcal{G}^{(e)}_{3'})(t,t')}^{\bar{t}}d(t-t')\,.
\end{equation}
As shown in appendix \ref{appendix:eh-count}, at vanishing temperature and for $\mu_{3'}=\mu_1$ this quantity is nothing but the average excess number
of excitations (electrons and holes) produced per period when the AC part of the drive is switched on. 
At non zero temperature, thermal electron and hole excitations in the second incoming channel antibunch with the electrons and
hole excitations emitted by the source of pulses thus leading to a reduction of the noise\cite{Bocquillon:2012-1}:
\begin{equation}
\label{eq:HBT:noise-signal}
\frac{\Delta S_{1',1'}^{\mathrm{exp}}}{\mathcal{R}\mathcal{T}}=\frac{e^2}{2\pi}
\int_{-\infty}^{+\infty} \mathrm{tanh}{\left(\frac{\hbar\omega}{k_BT_{\mathrm{el}}}\right)}
\,v_F\Delta \mathcal{G}^{(e)}_{1,0}(\omega)\,d\omega\,
\end{equation}
where one recognizes the excess occupation number generated by the ac drive: $v_F\Delta \mathcal{G}^{(e)}_{1,0}(\omega)=f_{\mathrm{ac}+\mathrm{dc}}(\omega)-f_{\mathrm{dc}}(\omega)$. Note that the excess noise reduction
due to the $\tanh{(\hbar\omega/k_BT_\mathrm{{el}})}$ factor  
concerns electron and hole excitations emitted close to the chemical potential $\mu_1$.

\medskip

Combining \eqref{eq:HBT:noise-signal} with the electron distribution function \eqref{eq:Floquet:population} leads to the following
expression for the excess noise in terms of the photo-assisted transition probabilities\cite{Dubois:2012-1}:
\begin{eqnarray}
\label{eq:HBT:excess-noise}
\frac{\Delta S_{1',1'}^{(\mathrm{exp})}}{\mathcal{RT}e^2f} & = & \sum_{l=-\infty}^{+\infty} 
P_l[V_{\mathrm{ac}}]\,(l+\alpha)\coth{\left(\frac{(l+\alpha)hf}{2k_BT_{\mathrm{el}}}\right)}\nonumber\\
 & - & \alpha \coth{\left(\frac{\alpha hf}{2k_BT_{\mathrm{el}}}\right)}\,.
\end{eqnarray}
In the limit of vanishing temperature, the excess noise $\Delta S_{1',1'}^{(\mathrm{exp})}$
is given in terms of the average number of electron and holes excitations $\bar{N}_e$ and $\bar{N}_h$ emitted per period by the source:
\begin{eqnarray}
\frac{\Delta S_{1',1'}^{\mathrm{exp}}}{\mathcal{R}\mathcal{T}} & = & e^2f\sum_{l=-\infty}^{+\infty} 
P_l[V_{\mathrm{ac}}]\,(|l+\alpha|-|\alpha|)\nonumber \\
& = & e^2f\left(
(\bar{N}_e+\bar{N}_h)_{T_{\mathrm{el}}=0}-|\alpha|\right)\,.
\end{eqnarray}
Our previous work\cite{Dubois:2012-1} presents a detailed study of this quantity at zero temperature for various
pulse shapes and in realistic temperature conditions. In the case of Lorenztian pulses, the zero temperature
excess noise vanishes when $\alpha$ is a positive (resp. negative) integer since the source then emits $\alpha$ electrons 
and no hole (resp. $-\alpha$ holes and no electron). For non integer values of $\alpha$, we observe a non zero electron/hole
pair production as expected from the physics of orthogonality catastrophe unraveled by Levitov {\it et al}\cite{Lee:1993-1,Levitov:1996-1}.

In the next section, we will consider the effect of interactions on voltage pulses and see 
to what extent shot noise provides a way to study charge fractionalization. As we shall see now, expression
\eqref{eq:HBT:excess-noise} can still be used in presence of interactions provided 
the photo-assisted probabilities are computed from the appropriate AC voltage.

\section{Decoherence and relaxation}
\label{sec:decoherence}

\subsection{Floquet approach to decoherence for a periodic train of pulses}
\label{sec:decoherence:general}

\subsubsection{Edge magnetoplasmon scattering}

To deal with interactions, it is convenient to use the bosonization framework which describes
all excitations of the 1D chiral electronic fluid in terms of edge magnetoplasmon modes $b(\omega)$
and $b^\dagger(\omega)$ ($\omega>0$) which are directly related to the finite frequency electrical
current: $i(\omega>0)=e\sqrt{\omega}\,b(\omega)$. 

We then consider an interacting region where electrons within the edge channel experience electron/electron interactions or 
Coulomb interactions with other conductors described as a linear environment. Due to linearity, interaction effects in this region lead
to an elastic scattering between the edge magnetoplasmon modes 
and the environmental modes. These environmental modes possibly include
the edge magnetoplasmon modes of another edge channel in the
case of a coupled $\nu=2$ edge channel system\cite{Degio:2010-1} as well as the electromagnetic modes
of another mesoscopic conductor\cite{Degio:2009-1}. Denoting by
$a(\omega)$ and $a^\dagger(\omega)$ where $\omega>0$ the destruction and creation 
operators of these modes, the effect of interactions are described by a scattering matrix $S(\omega)$ such that:
\begin{equation}
\label{eq:Levitov:plasmon-scattering}
\left(\begin{array}{c}
a_{\mathrm{out}}(\omega) \\
b_{\mathrm{out}}(\omega) 
\end{array}\right) = S(\omega)\,
\left(\begin{array}{c}
a_{\mathrm{in}}(\omega) \\
b_{\mathrm{in}}(\omega) 
\end{array}\right)\,.
\end{equation}
As in quantum wires\cite{Safi:1999-1,Safi:1995-2}, the magnetoplasmon scattering matrix is directly 
related to finite frequency admittance\cite{Degio:2009-1,Degio:2010-1}: $G_{ee}(\omega)=(e^2/h)(1-S_{bb}(\omega))$
where $S_{bb}(\omega)$ connects $b_{\mathrm{in}}(\omega)$ to $b_{\mathrm{out}}(\omega)$.

\medskip

As a consequence, it is possible to access the edge magnetoplasmon scattering amplitudes through a finite frequency
admittance measurement. In the case of the setup depicted on figure \ref{fig:HBT}, such measurements can be performed
by polarizing the first QPC so that it reflects both edge channels or only the inner one, thus giving access
to finite frequency admittance ratios. Such a measurement has recently been performed\cite{Bocquillon:2012-2} and,
for the first time, has lead to direct information on edge magnetoplasmon scattering which is
used to discuss electronic decoherence and relaxation in the $\nu=2$ edge channel 
system\cite{Levkivskyi:2008-1,Degio:2010-1,Kovrizhin:2009-1}. 

Knowing the finite frequency admittance,
we can  determine how interactions alter a periodic train of pulses.

\subsubsection{Floquet approach and interactions}
\label{sec:decoherence:Floquet}

At zero temperature, a classical time dependent voltage drive $V(t)$ generates
a coherent state for all magnetoplasmon modes above the shifted Fermi sea $|F_{\mu+\alpha hf}\rangle$. 
Introducing the notation $|[\Lambda(\omega)]\rangle$ for as coherent state characterized by 
the complex eigenvalues $\Lambda(\omega)$ for each $b(\omega)$ for $\omega>0$, and using the expression of the 
electric current in terms of $b(\omega)$ and $b^\dagger(\omega)$ as well as the current/voltage relation $\langle i(t)\rangle = (e^2/h)V(t)$,
these parameters are related to the voltage drive by
\begin{equation}
\label{eq:coherent:voltage-relation}
\Lambda(\omega)=-\frac{e\widetilde{V}(\omega)}{h\sqrt{\omega}}\,
\end{equation}
where $\widetilde{V}(\omega)=\int V(t)e^{i\omega t}dt$. Thus, for a periodic train of pulses, only the
modes at harmonic frequencies of the driving frequency $f$ are excited. At zero temperature, the incoming 
state for the environmental modes is the vacuum state with respect to
the environmental modes which we denote by $|[0_\omega]\rangle$. 

Therefore, the incoming
factorized state $|[\Lambda(\omega)]\rangle\otimes |[0_{\omega}]\rangle$ comes out from the interaction region 
as a factorized coherent state because the interaction region acts as a frequency dependent
beam splitter for the edge magnetoplasmon and environmental modes\footnote{A tensor product of coherent states scatters into a tensor 
product of coherent states whose parameters are inferred from the scattering matrix of the beam splitter. This reflects the way classical waves are transmitted through a beam splitter in the quantum theory.}.
Consequently, tracing out the environmental degrees of freedom leads to a coherent state for the edge magnetoplasmon modes. 
This reduced state for the edge
channel under consideration is of the form $|[S_{bb}(\omega)\Lambda(\omega)]\rangle$. 

\medskip

Therefore, the outcoming state is a pure (many-body) state generated by a distorted voltage pulse: $V_{\mathrm{out}}(\omega)=S_{bb}(\omega)
V_{\mathrm{in}}(\omega)$. As a consequence, the effect of interactions on a pulse generated state can always corrected
by applying an appropriate correction to the voltage pulse\cite{Lebedev:2011-1}.
In the case of a periodic train of pulses, it also implies that the outcoming 
single electron coherence $\mathcal{G}^{(e)}_{\mathrm{out}}$ can be computed 
using the Floquet approach presented in section \ref{sec:source:periodic}. 

This result does indeed extend to non zero electronic temperature $T_{\mathrm{el}}$ since then, the state generated by a classical time dependent
voltage drive is a displaced thermal state at temperature $T_{\mathrm{el}}$ and the environmental incoming state is the thermal equilibrium state
at the temperature of the environment. As discussed in appendix \ref{appendix:finite-temperature}, 
when both temperatures are equal, the resulting outcoming state is factorized into two displaced thermal states
at the common temperature. Moreover, their displacements are given by the classical scattering, exactly as in the zero temperature situation.
Consequently, the outcoming single electron coherence can still be computed within the framework of Floquet scattering theory\cite{Moskalets:book},
taking into account the non zero temperature of the electronic reservoirs.

\medskip

To summarize, we have shown that the effect of interactions on a periodic train of voltage pulses can be computed
exactly by combining the bosonization treatment of interactions and the Floquet  formalism. 
We shall now apply this method to discuss the case of the system of two coupled co-propagating edge
channels realized at filling fraction $\nu=2$.  

First, we shall consider the simplest model for
this system which assumes short range inter channel Coulomb interactions and leads to perfect
spin/charge separation which in turn induces an interesting stroboscopic revival of single electron coherence. 
However, in a real sample, screening may not be so efficient and therefore, we shall also consider
the opposite limit of long range interactions over a distance $l$ using the discrete element model originally introduced by Büttiker and 
his collaborators\cite{Buttiker:1993-1,Pretre:1996-1,Christen:1996-1}. Of course the method could be adapted to other models,
phenomenological\cite{Bocquillon:2012-2} or more microscopic\cite{Han:1997-2} that include dissipation of edge magnetoplasmon
modes but for simplicity we will focus on the two aforementioned models.

Finally, let us point out that the same approach could be used to discuss the effect
of interactions in a system at filling fraction $\nu=1$. However, in such a system, interaction within the edge channel lead to electronic relaxation and decoherence
if and only if they induce a frequency dependent time delay\cite{Neuenhahn:2008-1,Neuenhahn:2009-1}. This is the case as soon as we consider finite range interactions. 
This means that the edge magnetoplasmons of the $\nu=1$ edge channel system are dispersive and consequently that the shape of voltage voltage pulses is altered. In fact, $\nu=2$ is the first integer filling fraction which can lead to non trivial interaction effects without edge magnetoplasmon dispersion.

\subsection{Stroboscopic coherence revivals}
\label{sec:revivals}

In the case of a $\nu=2$ edge channel system with short range screened Coulomb interactions, the 
plasmon scattering matrix associated with an interaction region of length $l$ 
reflects the existence of two dispersionless magnetoplasmon modes delocalized on both edges within the interaction region. 
It has the following form\cite{Degio:2010-1}:
\begin{equation}
\label{eq:two-channels:universal-scattering}
S(\omega,l)=e^{i\omega l/v_0}\,e^{-i\omega (l/v)(\cos{(\theta)}\sigma^z+\sin{(\theta)}\sigma^x)}\,
\end{equation}
where $v_{0}$ and $v$ are two velocities and $\theta$ is an angle encoding the interaction strentgh. Uncoupled channels correspond to $\theta=0$ whereas
strongly coupled channels correspond to $\theta=\pi/2$. In the latter case, the eigenmodes are the symmetric and antisymmetric
linear combinations of the two channel magnetoplasmon modes respectively called the charge and dipolar modes\cite{Levkivskyi:2008-1,Levkivskyi:2012-1}.
The magnetoplasmon eigenmodes velocities are $v_{\pm}=v_{0}v/(v_{0}\pm v)$ and the charge mode is faster than the dipolar mode. 
Due to the absence of dispersion of these eigenmodes, each current pulse generated by the source splits into 
two pulses of the same shape and of respective weights $(1\pm \cos{(\theta)})/2$:
\begin{equation}
V_{1,\mathrm{out}}(t)=\frac{1+\cos{(\theta)}}{2}\,V(t-l/v_{+})+\frac{1-\cos{(\theta)}}{2}\,V(t-l/v_{-})\,.
\end{equation}
This is the fractionalization of a charge $-\alpha e$ pulse. In particular, at strong coupling ($\theta=\pi/2$),
we expect a single electron pulse to fractionalize into two Lorentzian pulses with $\alpha=1/2$ which can only be described
in terms of collective electron/hole pair excitations.

\medskip

Note that, in the case of
a periodic train of pulses, an interesting phenomenon occurs when the two pulses issued from consecutive pulses recombine. 
When $l$ is an integer multiple of $l_r=vT/2$, the original current pulse is restored with a half period shift. 
This phenomenon can be viewed as a stroboscopic revival of the original train of excitations. 

\begin{figure}
\includegraphics[width=8cm]{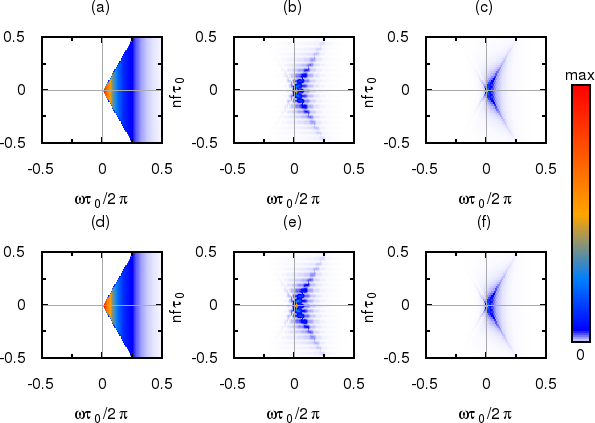} 
\caption{\label{fig:dispersionless:frequency} (Color online) Modulus of the single electron coherence function in the frequency domain emitted by 
a periodic source of single electron pulses ($\alpha=1$) with $\tau_{0}f=0.01$ as a function of $\omega\tau_{0}$ and $nf\tau_{0}$ in the 
presence of interactions at
strong coupling ($\theta=\pi/2$) and for various propagation lengths: (a) $l/l_{r}=0$, (b) $l/l_{r}=0.2$, (c) $l/l_{r}=0.5$,
(d) $l/l_{r}=1$, (e) $l/l_{r}=1.2$ and (f) $l/l_{r}=1.5$.}
\end{figure}

Figure \ref{fig:dispersionless:frequency} depicts the single electron coherence in 
the frequency domain at strong coupling ($\theta=\pi/2$) and for increasing propagation
lengths when a periodic train of single electron pulses is injected within one edge channel. 
The stroboscopic revival is clearly visible as well as the appearance of hole excitations and of electron/hole coherences for propagation lengths
between two revivals. Note the periodicity of the single electron coherence in $l\rightarrow l+l_r$ which arises from the 
dispersionless character of the edge magnetoplasmon eigenmodes within the interaction region, as can be seen
from the explicit form of the scattering matrix \eqref{eq:two-channels:universal-scattering}. 

In section \ref{sec:HBT}, we will present
predictions for the excess noise and electron/hole pair production in the $\nu=2$ edge channel system and discuss
the signature of this fractionalization phenomenon both in the case of short and long range interactions.

\subsection{The discrete element model}
\label{sec:HBT:long-range}

To investigate the effect of long range interactions, we have used a discrete element approach to the two channel system. 
In this approach, pioneered by B\"uttiker and his collaborators\cite{Buttiker:1993-1,Pretre:1996-1,Christen:1996-1} , each 
copropagating edge channel $j$ within the interacting region $0\leq x\leq l$  experiences a uniform time dependent potential $U_{j}(t)$. These potentials are 
related to the charges stored on each edge channel within the interacting region by a $2\times 2$ capacitance matrix. This model 
is expected to describe the physics of the two channel system when interactions are unscreened
and at low enough energy so that excitations sent into the interaction region do not feel the inhomogeneities of the electrostatic potential.

\medskip

As shown in appendix \ref{appendix:discrete-elements}, the scattering matrix for the edge magnetoplasmons can be computed analytically 
by solving the equations of motions for the corresponding bosonic fields. As in the short range case, the scattering reflects the existence of
eigenmodes which are linear combinations of the two edge magnetoplasmon modes. In the present case, their mixing angle $\theta$ is given by
\begin{equation}
\label{eq:two-channels:cos-sin}
\exp{(i\theta)}  =  \frac{\Delta C/2+i\,C}{\sqrt{ C^2 + \frac{\Delta C^2}{4}}}\,.
\end{equation}
where $\Delta C=C_1-C_2$ denotes the difference of the diagonal terms of the capacitance matrix and $C$ is the off diagonal term representing the interchannel
capacitance. Strong coupling is realized as soon as $C\gg |\Delta C|$. Long range interactions lead to dispersive propagation of these 
eigenmodes along a distance $l$. This has an important
consequence for all voltage pulses: after some copropagation, their shape is not preserved. 
As we shall see now, this has important consequences on the observability of the fractionalization phenomenon in
the $\nu=2$ quantum Hall edge channel.

\section{Probing fractionalization through Hanbury Brown and Twiss interferometry}
\label{sec:HBT}

To study fractionalization in the $\nu=2$ edge channel system, we consider the electron/hole
pair production as a function of $\alpha$ and of the copropagation distance $l$
which can be accessed via low frequency current noise measurements in the Hanbury Brown and Twiss
geometry as explain in \ref{sec:characterization:hbt}.  As we shall see now, the dependence 
of electron/hole production in $(\alpha,l)$ contains clear signatures of fractionalization and also qualitative features 
enabling to distinguish between short and long range interactions. 

\subsection{Fractionalization in the $\nu=2$ edge channel system}
\label{sec:HBT:fractionalization}

\subsubsection{Short range interactions}
\label{sec:HBT:fractionalization:short-range}

For short range interactions, a Lorentzian pulse of charge $-\alpha e$ splits into two Lorentzian pulses of respective charges $-\alpha e\cos^{2}{(\theta/2)}$ and 
$-\alpha e\sin^{2}(\theta/2)$. When the charges of these pulses are non integer such as for $\alpha=1$ (single electron pulses), 
a non vanishing electron/hole pair production is expected at a generic distance ({\it i.e.} not a multiple of the revival distance $l_{r}$).

\medskip

Nevertheless, electron/hole pair production does vanish when 
fractionalization leads to purely electronic pulses. For short range interactions, this is the case when $\alpha\cos^{2}{(\theta/2)}$ and $\alpha\sin^{2}{(\theta/2)}$ 
are both integers. This first implies
that $\alpha$ is an integer and, consequently, that $\cos^{2}{(\theta/2)}$ is a rational number. Assuming that $\cos^{2}{(\theta/2)}=p/p'$ where 
$p$ and $p'$ are two mutually prime numbers,
as soon as $\alpha$ is a multiple of $p'$, electron/hole pair production will vanish for any co-propagation length. These lines of vanishing
electron/hole pair production in the $(\alpha,l/l_r)$ plane correspond to the fractionalization of the charge $-\alpha e$ in two integer charges and therefore 
we shall call them {\em integer fractionalization lines}. 

\medskip

\begin{figure}
\includegraphics[width=9cm]{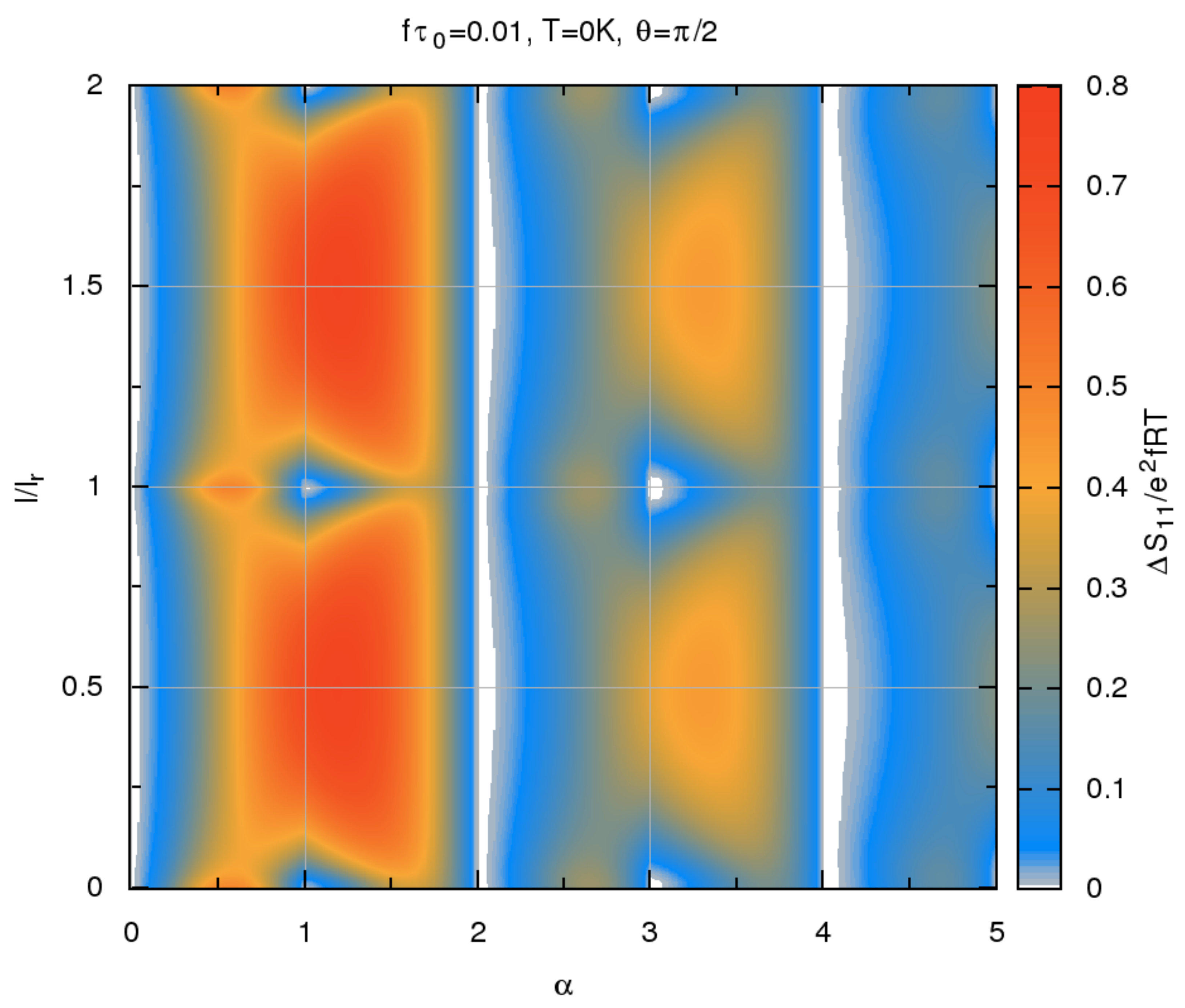}
\caption{\label{fig:ehp-colorplot-strong-coupling} (Color online) Density plot representing the
excess noise at zero temperature in the $(\alpha,l/l_r)$ plane for Lorentzian pulses with $f\tau_0=0.01$ in the presence of short range interactions
in the strong coupling regime ($\theta=\pi/2$). The integer fractionalization line at $\alpha=2$ and
$\alpha=4$ are clearly visible as white vertical zones. 
Note also the excess noise minima (white spots) around $(\alpha,l/l_r)=(1,n)$, $(3,n)$ for $n=0$, $1$ and $2$. 
}
\end{figure}

The density plot of figure \ref{fig:ehp-colorplot-strong-coupling} presents the excess noise for Lorentzian pulses at zero temperature in the
$(\alpha,l/l_r)$ plane at strong coupling ($\theta=\pi/2$) and for $f\tau_{0}=0.01$. 
It clearly shows that the $\alpha=2$ and $\alpha=4$ Lorentzian pulses are always fractionalized into 
purely electronic pulses for any $l/l_r$. We also see that when $l/l_r$ is close to an integer, the excess noise also decreases for $\alpha=1$
and $\alpha=3$. At a higher value of $f\tau_0$, the same qualitative picture can still be observed but the peak values for
the excess noise in units of $e^2f$ are lower, thus making it more difficult to observe. This decrease, specific to Lorentzian pulses\cite{Dubois:2012-1},
is indeed related to the Pauli exclusion principle and reflect the fact that 
when $\alpha$ increases from a positive integer $n$ to $n+1/2$, the 'half-electronic' excitation is
added on top of a many-body state obtained by adding on top of the Fermi sea the $n$-particle state discussed in section \ref{sec:source:fermionization} 
on top of the Fermi surface. Single particle states just above the Fermi energy being more and more occupied with increasing $n$, the 
number of hole excitations generated decreases with $n$. Nevertheless, 
observing this pattern of minimas for the excess noise in the $(\alpha,l/l_r)$ plane would be a clear signature of fractionalization
in the $\nu=2$ edge channel system. Before analyzing in more detail the effect of the pulse shape and temperature which are
relevant for the experiments, let us discuss the position of these fractionalization lines.

\begin{figure}
\includegraphics[width=7cm]{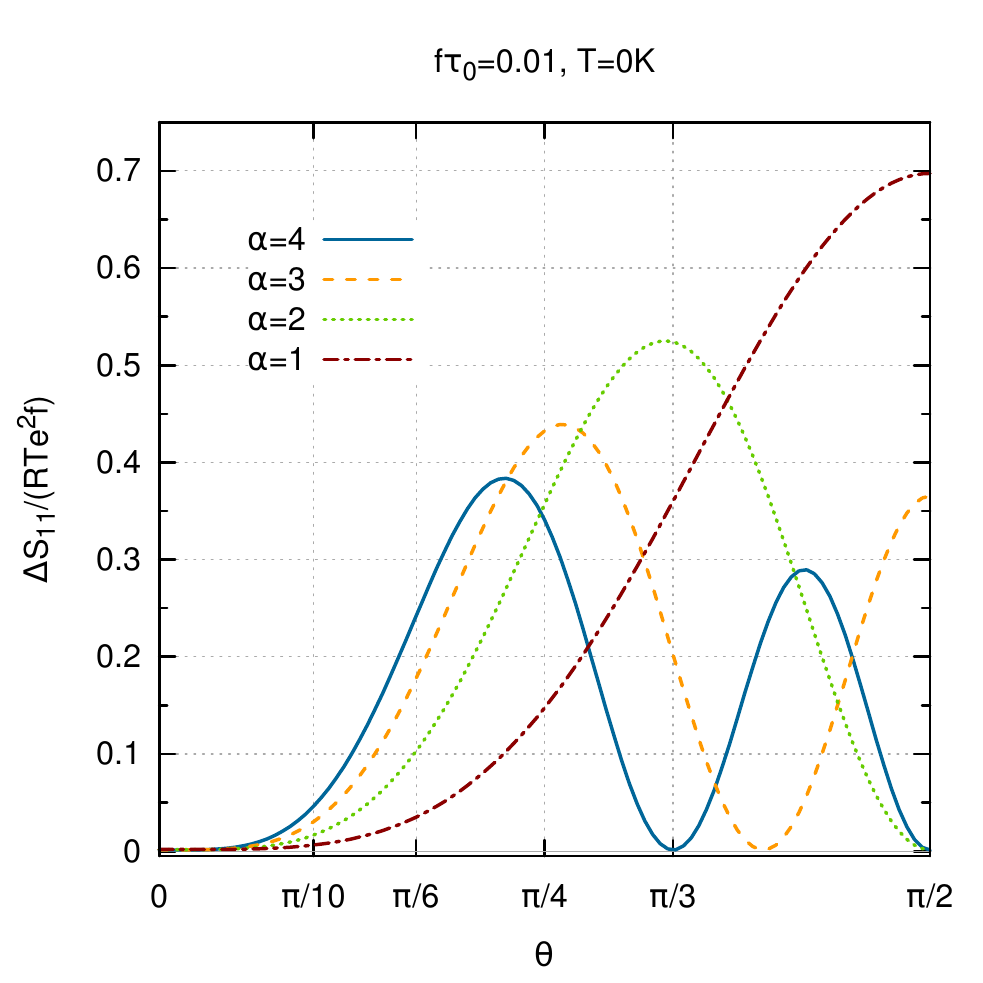}
\caption{\label{fig:theta-effect} (Color online) Electron/hole pair production in the presence of 
short range interactions, observed at $l=l_{r}/2$ as a function of
$\theta$ for $\alpha=1$ to $\alpha=4$ Lorentzian pulses with $f\tau_0=0.1$.
The vanishing of excess noise for these various curves occurs at the angles listed in table \ref{table:fractionalization}. Note 
that excess noise decreases with increasing $\alpha$.}
\end{figure}

Integer fractionalization lines are determined by the interaction strength and therefore by the angle $\theta$. 
Figure \ref{fig:theta-effect} represents the hole production 
computed at $l=l_r/2$ as a function of $\theta$ for $\alpha=1$ to $\alpha=4$ Lorentzian pulses. As expected,
the $\alpha=1$ pulse never leads to a vanishing electron/hole pair production as soon as $\theta\neq 0$. The $\alpha=2$ Lorentzian
pulses lead to vanishing electron/hole production when it is split into two pulses of equal amplitudes: this occurs at $\theta=\pi/2$. For $\alpha=3$ Lorentzian 
pulse, splitting into charge $2$ and $1$ pulses occurs when $\cos{(\theta)}=\pm 1/3$. The $\alpha=4$ case leads to a vanishing 
of electron/hole pair production for $\theta=\pi/3$ corresponding to a splitting $4=3+1$ of the charge, and $\theta=\pi/2$ corresponding to $4=2+2$. 
These integer fractionalization lines are listed in table \ref{table:fractionalization} for 
$2\leq \alpha \leq 4$. Finding these lines gives a rough estimation of $\theta$ as shown on 
figure \ref{fig:theta-effect}. However, let us stress that finite frequency admittance measurements\cite{Bocquillon:2012-2}
give an independent and more precise determination of the angle $\theta$. Combining these measurements would thus provide a consistency
check of our edge magnetoplasmon scattering approach to relaxation and decoherence of pulse excitations.

\begin{table}
\begin{center}
\begin{tabular}{|c|c|c|c|c|}\hline
$\alpha$ & $\theta$ & $\cos^2{(\theta/2)}$ & $\sin^2{(\theta/2)}$ \\ \hline\hline
$1$ & $0$ & $1$ & $0$ \\  \hline
$2$ & $\pi/2$ & $1/2$ & $1/2$ \\ \hline
$3$ & $\arccos{(1/3)}$ & $2/3$ & $1/3$ \\
 & $\arccos{(-1/3)}$ & $1/3$ & $2/3$ \\ \hline
$4$ & $\pi/3$ & $3/4$ & $1/4$ \\
	& $\pi/2$ & $2/4$ & $2/4$ \\
 & $2\pi/3$ & $1/4$ & $3/4$ \\ \hline
\end{tabular}
\end{center}
\caption{\label{table:fractionalization} Integer fractionalization lines for $2\leq \alpha\leq 4$ Lorentzian pulses. For each
value of $\alpha$, the values of $\theta$ for which integer fractionalization of $\alpha$ pulses take place and the corresponding fractions are given.}
\end{table}

As can be seen from figure \ref{fig:ehp-colorplot-strong-coupling}, increasing $\alpha$ does not improve the observability of fractionalization
as could be expected from our previous work\cite{Dubois:2012-1}:
since local interactions split pulses into pulses of the same shape, we expect the effect of interaction induced 
fractionalization to be lower at higher $\alpha$ for Lorentzian pulses. 

\subsubsection{Finite range interactions}
\label{sec:HBT:fractionalization:long-range}

Finite range interactions lead to the distortion of the injected pulses and therefore will affect the electron/hole
pair production. A natural question is therefore to understand how finite range interactions will affect the simple
physical picture discussed in the previous paragraph. Although the answer to this question is model dependent, the discrete element 
model provides an answer in the case of long range interactions between the two edge channels. 

\medskip

\begin{figure}
\includegraphics[width=7cm]{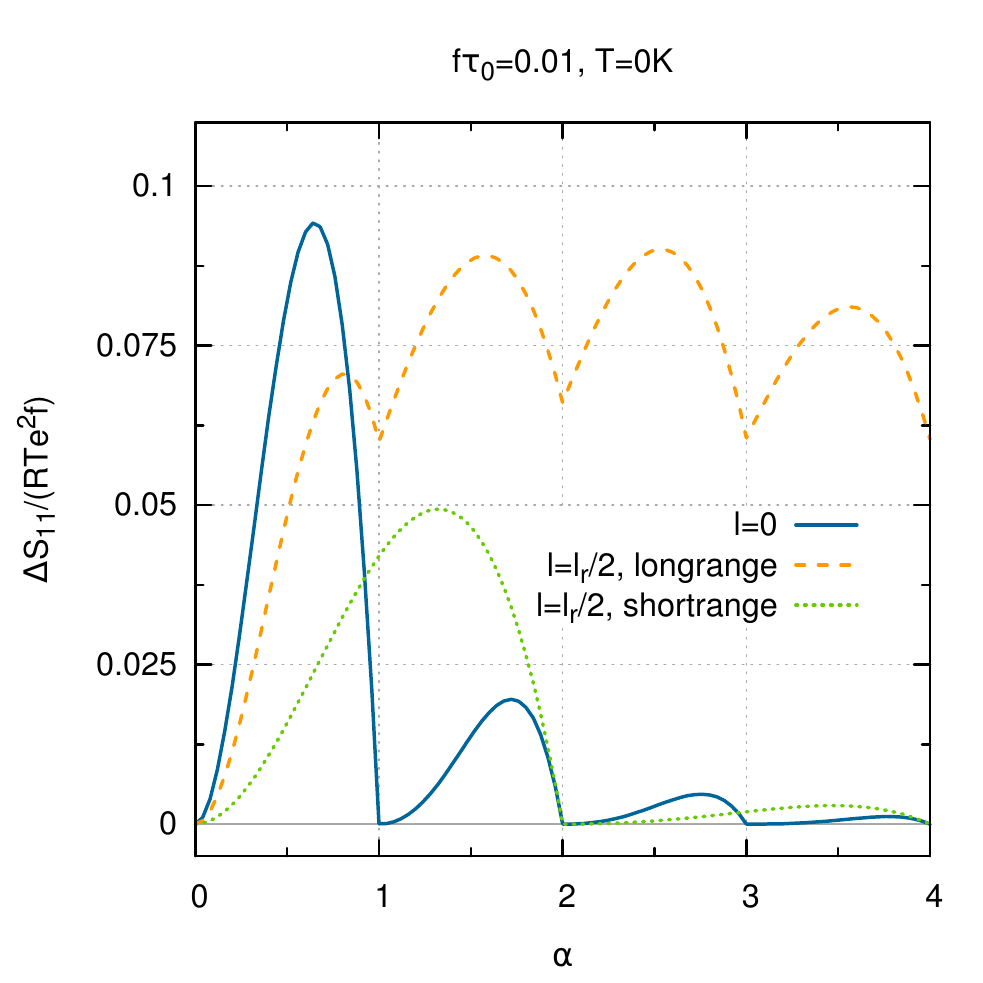}
\caption{\label{fig:interaction-range:ideal} (Color online) Electron/hole pair production in the presence of 
long and short range interactions at strong coupling, observed at $l=l_{r}/2$ as a function of
$\alpha$ for Lorentzian pulses with $f\tau_0=0.01$. The dashed curve correspond to long range interactions
and shows to minima at integer values of $\alpha$ with non vanishing electron/hole pair production. The full line correspond
to the non-interacting case obtained for $l/l_r=0$.}
\end{figure}

Figure \ref{fig:interaction-range:ideal} presents the zero temperature excess noise $\Delta S_{1,1}^{(\mathrm{out})}$
for Lorentzian pulses in the non interacting case compared to the predictions for the short range interaction model and
for the discrete element model. As expected, we observe vanishings
of the zero temperature excess noise for short range interactions whereas long range interactions leads to non vanishing minima. 
However, this strong quantitative difference is specific to the zero temperature case where the excess noise
exactly reflects the intrinsic electron/hole pair production of the excitations present within the system. But at finite
temperature, this signature of long range interactions may not be so clear. 

For this reason, it is important to consider more realistic situations taking into account the finite electronic temperature
as well as the non ideality of the pulses. In the forthcoming section, we will present numerical results for the excess noise
taking into account finite temperature for Lorentzian and rectangular pulses with realistic parameters. First we will discuss
the observability of fractionalization induced by short range interactions and next, we will consider the case of long range interactions.

\subsection{Numerical results}
\label{sec:HBT:numerics}

In this section, we present numerical results for the excess noise $\Delta S_{1,1}^{(\mathrm{out})}$ in units of $e^2f$ obtained by
applying the Floquet approach described in section \ref{sec:decoherence:Floquet} in the short range interaction model as well as in the
discrete element model. Both are considered at strong coupling ($\theta=\pi/2$), as expected from recent 
experiments\cite{Degio:2010-1,LeSueur:2010-1,Bocquillon:2012-2}.

We consider Lorentzian pulses with $\tau_0=120~\mathrm{ps}$ pulses at
a $1.2\ \mathrm{Ghz}$ drive frequency which could be generated by state of the art arbitrary waveform
generators (AWG). For rectangular pulses, we consider $\tau_0=40~\mathrm{ps}$ pulses since these can be generated 
with a single step by the AWG and therefore have the shortest available time step. In this case, $e^2f\simeq 3\times 10^{-29}\ \mathrm{A}^2/\mathrm{Hz}$. Our numerical results show that a sensitivity of the order of a percent of $e^2f$ is requested corresponding to a few $10^{-31}\ \mathrm{A}^2/\mathrm{Hz}$. Note that 
a $10^{-30}\ \mathrm{A}^2/\mathrm{Hz}$ sensitivity has been reached in recent experiments\cite{Parmentier:2010-1,Bocquillon:2012-1}.

Electronic temperatures $T_{\mathrm{el}}=5$, $10$, $20$ and $40$~mK have been considered to analyze the effect
of a temperature, keeping in mind that only the last one corresponds to realist experimental conditions. 
Note also that for $f=1.2\ \mathrm{GHz}$ and $T_{\mathrm{el}}=40\ \mathrm{mK}$, we are not
far from the transition to a classical behavior for the excess noise since $k_BT_{\mathrm{el}}/hf=0.67$. As shown before\cite{Dubois:2012-1}, the
oscillations of the excess noise with $\alpha$ reflecting the electron/hole content of the pulses 
can only be observed for $k_BT_{\mathrm{el}}\lesssim \hbar f$ which puts a rather strong constraint on the
electronic temperature.

\subsubsection{Finite temperature}
\label{sec:HBT:numerics:temperature}

\begin{figure}
\includegraphics[width=8cm]{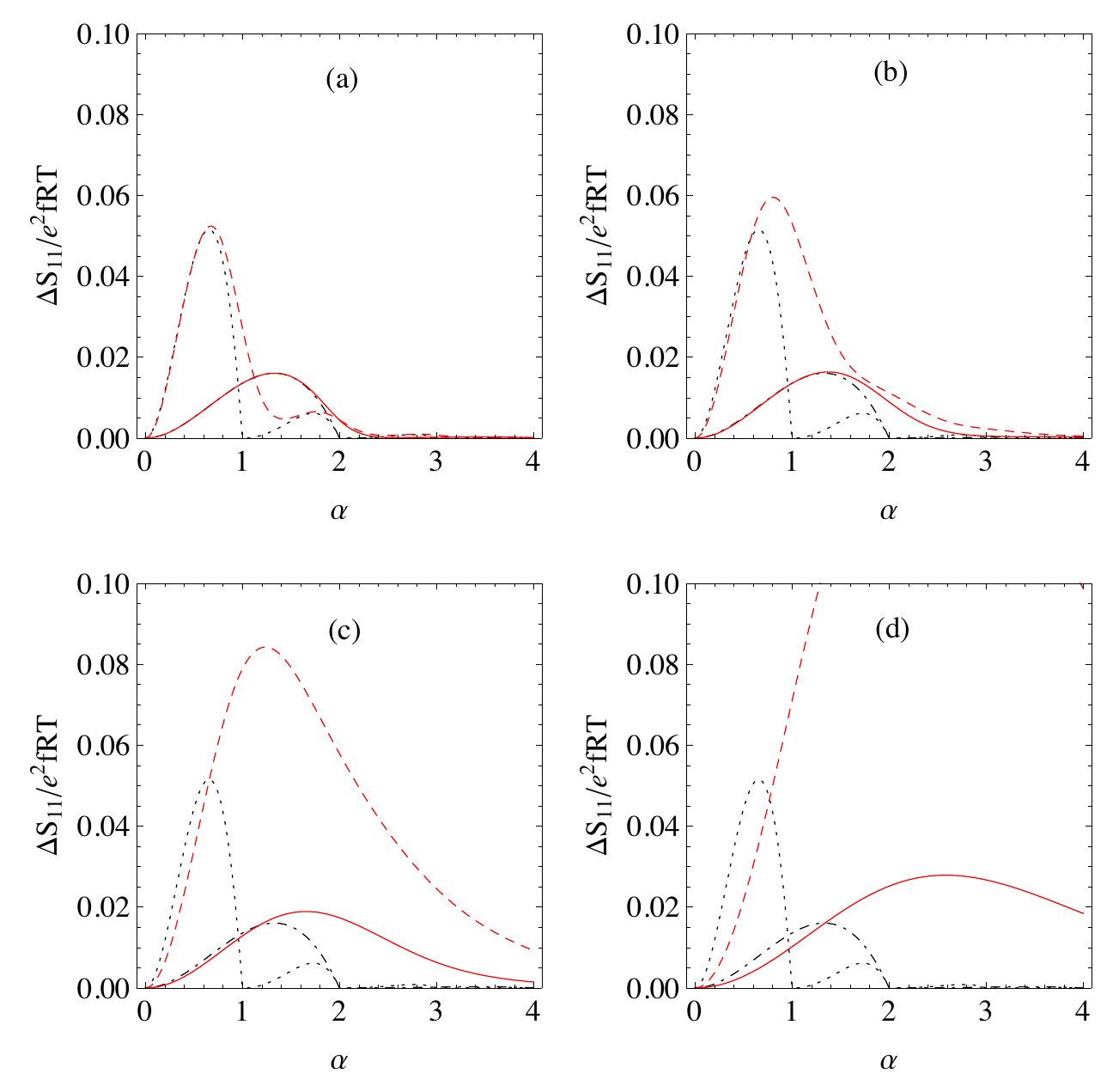}
\caption{\label{fig:interaction:ideal:temperature} (Color online) Excess noise for Lorentzian pulses of width $\tau_0=120\ \mathrm{ps}$ at
1.2~GHz ($f\tau_0=0.144$) in the presence
of short range interactions  in the strong coupling regime, observed at $l/l_r$ integer (red dashed curves) and $l/l_r$ half-integer (red full curves) as a function of
$\alpha$ for various temperatures: (a) $T_{\mathrm{el}}=5\ \mathrm{mK}$,  (b) $T_{\mathrm{el}}=10\ \mathrm{mK}$, (c) $T_{\mathrm{el}}=20\ \mathrm{mK}$
and (d) $T_{\mathrm{el}}=40\ \mathrm{mK}$. For comparison, we have replotted the zero temperature results on each panel 
as dotted and dot-dashed black curves.}
\end{figure}

The effect of a finite temperature on a periodic train of Lorentzian pulses in the presence of short range interactions 
are depicted on figure \ref{fig:interaction:ideal:temperature} where 
the excess noise is plotted as a function of $\alpha$ for the parameters given just above. It immediately shows that observing fractionalization with 
Lorentzian pulses might be very difficult. 

For 120~ps
pulses repeated at 1.2~GHz, observing the signature of fractionalization given by the position of excess noise minima
requires that $T_{\mathrm{el}}\lesssim 5\ \mathrm{mK}$ or, in dimensionless terms, $k_BT_{\mathrm{el}}/hf\lesssim1/12$ at $f\tau_0=0.144$. 
Reaching $f\tau_0=0.1$ and $k_BT_{\mathrm{el}}/hf\lesssim 0.1$ for an electronic temperature of 40~mK requires generating 10~ps Lorentzian pulses
at a repetition rate of 10~GHz. In these conditions, the fractionalization signature on the positions of excess noise minima would appear very clearly.
This could be achieved in the near future using optical methods\cite{Kato:2005-1}. 

\subsubsection{Changing the waveform}
\label{sec:HBT:numerics:waveform}

However, as can be noticed from figure \ref{fig:interaction:ideal:temperature}, part of the difficulty arises from the fact that
the difference between the minimal and maximal values of excess noise for Lorentzian pulses at zero temperature is
not very large and decays with $\alpha$ thus making the effect very fragile against thermal fluctuations. Since this is not the case for rectangular
pulses\cite{Dubois:2012-1}, a strategy for observing fractionalization might be to consider a rectangular waveform. 

\medskip

\begin{figure}
\includegraphics[width=8cm]{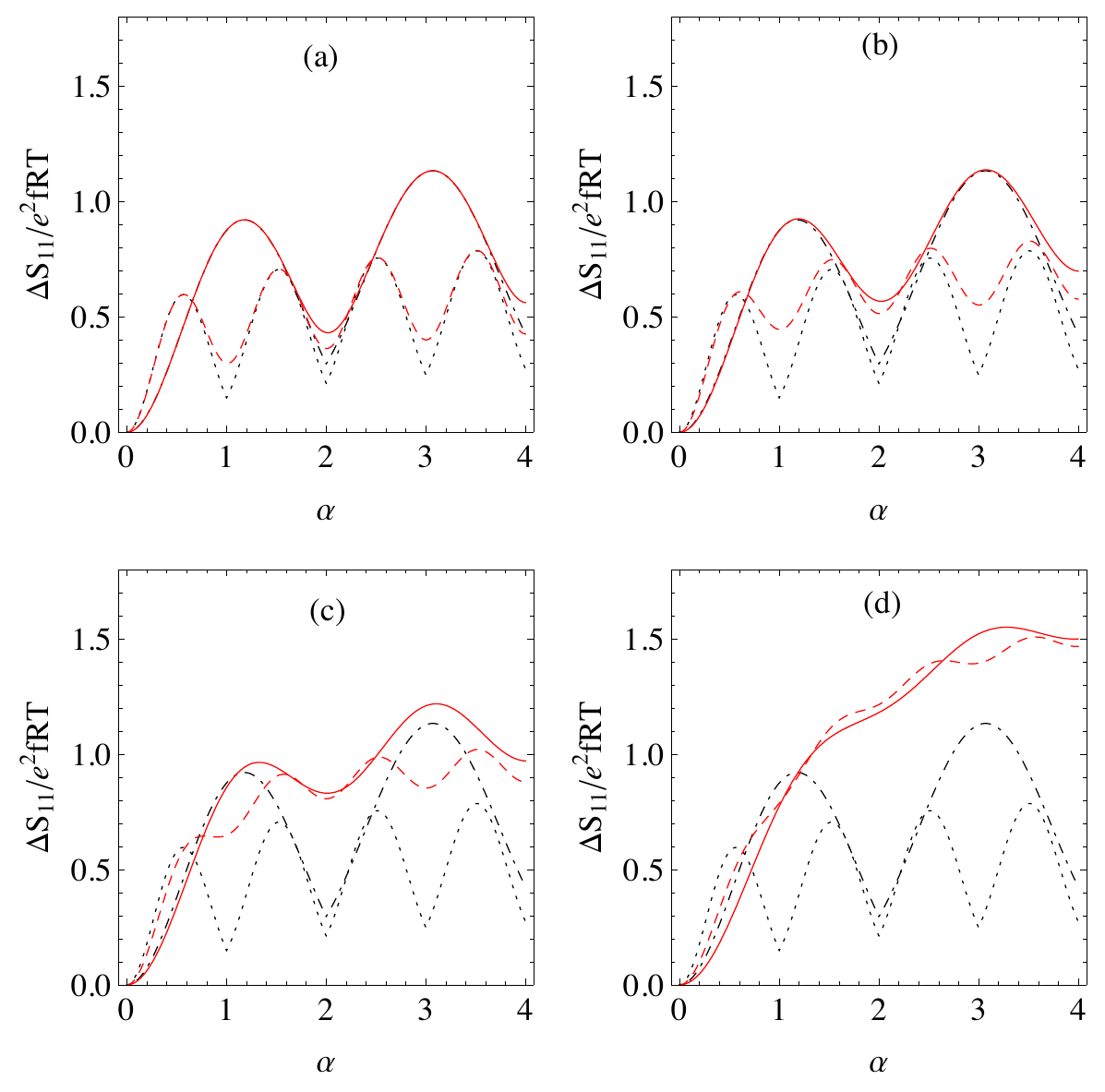}
\caption{\label{fig:interaction:square:temperature} (Color online) Excess noise for rectangular pulses of width $\tau_0=40\ \mathrm{ps}$ at
1.2~GHz ($f\tau_0=0.048$) in the presence
of short range interactions in the strong coupling regime, observed at $l/l_r$ integer (red dashed curves) and $l/l_r$ half-integer (red full curves) as a function of
$\alpha$ for various temperatures: (a) $T_{\mathrm{el}}=5\ \mathrm{mK}$,  (b) $T_{\mathrm{el}}=10\ \mathrm{mK}$, (c) $T_{\mathrm{el}}=20\ \mathrm{mK}$
and (d) $T_{\mathrm{el}}=40\ \mathrm{mK}$. For comparison, we have replotted the zero temperature results on each panel 
as dotted and dot-dashed black curves.}
\end{figure}

We have thus considered rectangular pulses for which, as mentioned before, shorter pulses could then be generated by an AWG. 
Figure \ref{fig:interaction:square:temperature}
corresponds $\tau_0=40~\mathrm{ps}$ pulses at the same $1.2$~GHz driving frequency and for the same temperatures as
in figure \ref{fig:interaction:ideal:temperature}. Interactions are assumed to be short range. 
As expected, the signal is stronger and more robust to the onset of the temperature.
Even at 20~mK, the position of the minima for the excess noise exhibit a clear signature of
the fractionalization of each $\alpha$ pulse into two $\alpha/2$ pulses. Therefore, although rectangular pulses with integer $\alpha$ are
not purely electronic excitations, they indeed appear as good candidates to demonstrate fractionalization at the single pulse level with today's
technology.

\subsubsection{Long range interactions}

\begin{figure}
\includegraphics[width=8cm]{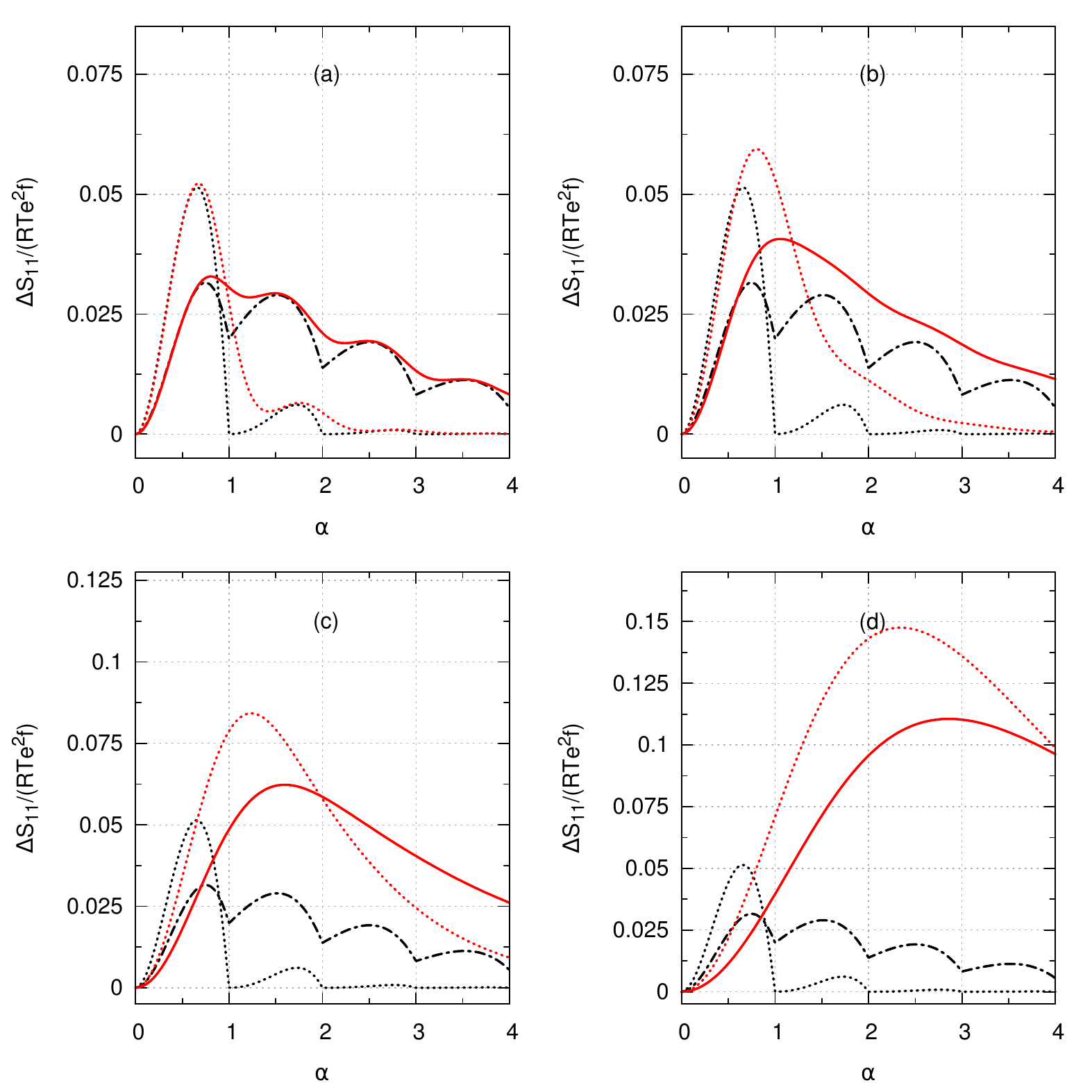}
\caption{\label{fig:interaction:dispersive:temperature} (Color online) Excess noise for Lorentzian pulses of width $\tau_0=40\ \mathrm{ps}$ at
1.2~GHz ($f\tau_0=0.048$) in the presence
of long range interactions  in the strong coupling regime, observed at $l/l_r$ integer (red dashed curves) and $l/l_r$ half-integer (red full curves) as a function of
$\alpha$ for various temperatures: (a) $T_{\mathrm{el}}=5\ \mathrm{mK}$,  (b) $T_{\mathrm{el}}=10\ \mathrm{mK}$, (c) $T_{\mathrm{el}}=20\ \mathrm{mK}$
and (d) $T_{\mathrm{el}}=40\ \mathrm{mK}$. For comparison, we have replotted the zero temperature results on each panel 
as dotted and dot-dashed black curves.}
\end{figure}

We now consider the case of long range interactions modeled using the discrete element model introduced 
in section \ref{sec:HBT:fractionalization:long-range}. Figure
\ref{fig:interaction:dispersive:temperature} present the case of Lorentzian pulses in the presence of long range
interactions with the same parameters as in figure \ref{fig:interaction:ideal:temperature}. Interaction parameters have
been chosen as described in appendix \ref{appendix:discrete-elements}: we have considered the strong coupling 
regime $\theta=\pi/2$ and at low frequencies, have estimated the charge velocity as $v_-\simeq 7.8\times 10^6\ \mathrm{ms}^{-1}$ 
and the spin velocity as $v_+\simeq 4\times 10^5\ \mathrm{ms}^{-1}$ which are compatible with known expected values
from time of flight\cite{Kamata:2010-1,Kumada:2011-1} and finite frequency admittance\cite{Bocquillon:2012-1} measurements.

\medskip

At the lowest temperature (graph (a)), the excess noise does not
exhibit for a generic length a pronounced minimum at $\alpha=2$ but instead starts decaying with very smooth local minima
slightly above each integer value of $\alpha$. Therefore, the signature of fractionalization discussed previously is
absent as expected since dispersion strongly alters the shape of the pulses. This induces a much
slower decay of the excess noise as a function of $\alpha$ for $l=l_r/2$ than in the case of short range interactions. This
could be used as a clear signature of the long range character of interactions at very low temperatures. 
Note that at higher temperature (graphs (d): $T_{\mathrm{el}}=40\ \mathrm{mK}$), only a quantitative comparison between experimental
data and theory would enable discriminating between various interaction models. 

\medskip

\begin{figure}
\includegraphics[width=8cm]{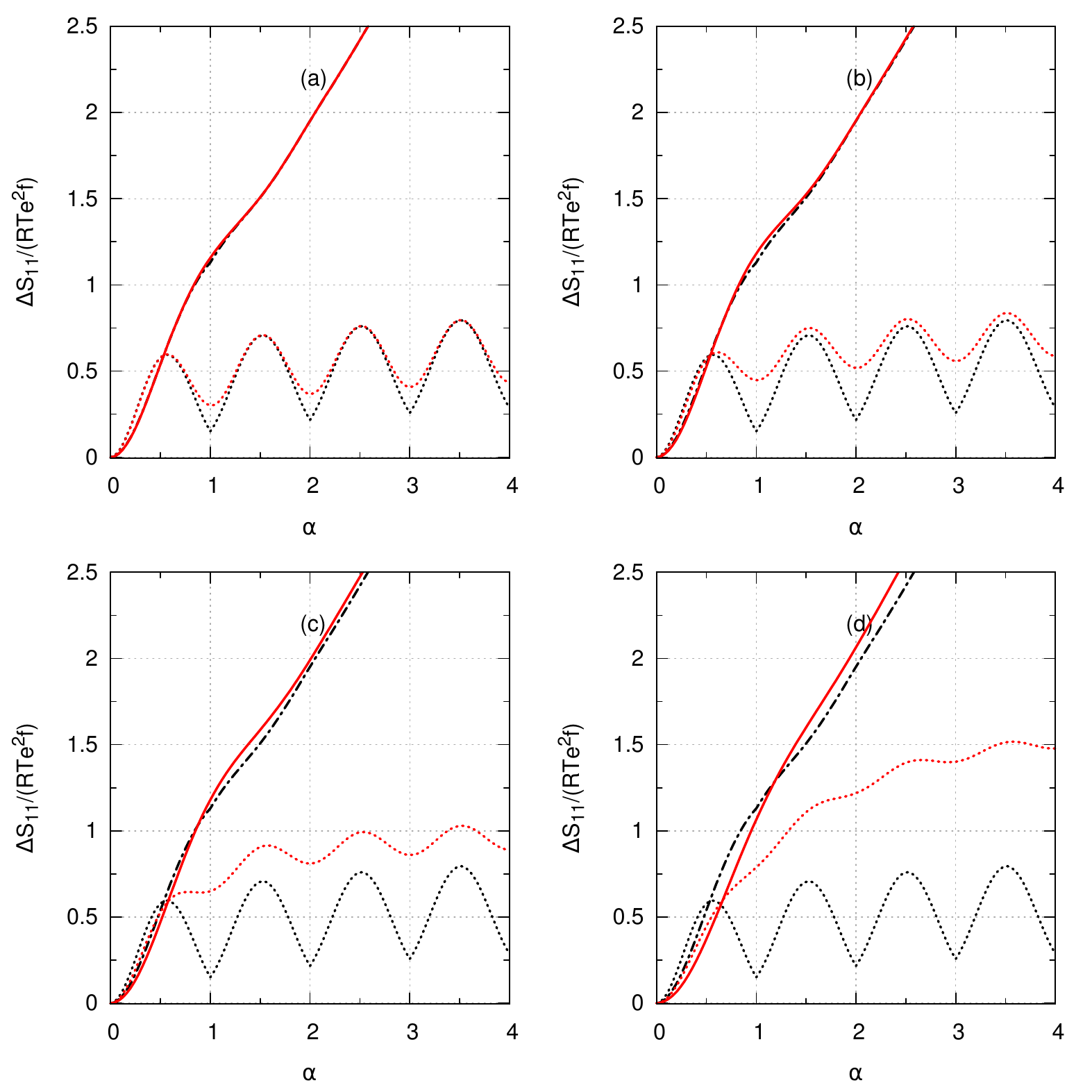}
\caption{\label{fig:interaction:dispersive:square::temperature} (Color online) Excess noise for rectangular 
pulses of width $\tau_0=120\ \mathrm{ps}$ at
1.2~GHz ($f\tau_0=0.144$) in the presence
of long range interactions  in the strong coupling regime, observed at $l/l_r$ integer (red dashed curves) and $l/l_r$ half-integer (red full curves) as a function of
$\alpha$ for various temperatures: (a) $T_{\mathrm{el}}=5\ \mathrm{mK}$,  (b) $T_{\mathrm{el}}=10\ \mathrm{mK}$, (c) $T_{\mathrm{el}}=20\ \mathrm{mK}$
and (d) $T_{\mathrm{el}}=40\ \mathrm{mK}$. For comparison, we have replotted the zero temperature results on each panel 
as dotted and dot-dashed black curves.}
\end{figure}

Considering rectangular waveforms does not radically change the picture: at very low temperature, the difference between short and long
range interactions is very clear, this time even more qualitative since for short range interactions, minima of the excess noise in function
of $\alpha$ occur for the three lowest temperatures considered (graph (a), (b) and (c) of figure \ref{fig:interaction:square:temperature}) 
whereas, in the case of long range interactions
the excess noise increases rapidly with $\alpha$ (graph (a), (b), (c) and (d) of figure \ref{fig:interaction:dispersive:square::temperature}). This
is qualitatively and quantitatively different from predictions obtained from the short range model.

\section{Conclusion}
\label{sec:conclusion}

In a system of two copropagating edge channels, interactions lead to charge fractionalization. We propose to detect
this phenomenon at the single excitation level by monitoring the electron/hole pair production of voltage pulses as a function
of their charge. More precisely, electron/hole pair production presents a minima 
each time an integer charge pulse is fractionalized into two pulses of integer charges.  

To discuss the observability of this phenomenon in realistic experiments, we 
have studied  the excess noise coming out of  and HBT interferometer fed by a periodic train or Lorentzian or rectangular
pulses. Although experimentally challenging, our study suggests that Lorentzian 
pulses of short enough durations could also be used to probe in detail the physics of fractionalization:
beyond the characterization of electron/hole pair production, current noise measurement in 
the Hanbury Brown and Twiss  setup can be used to perform a lest a spectroscopy\cite{Dubois:2012-1} and, ultimately, 
a full tomography\cite{Degio:2010-4} of fractional charge collective excitations. We have also shown that fractionalization of rectangular pulses could 
be observed with state of the art microwave waveform generators and that, for Lorentzian pulses, it might be within reach in the near future using optical techniques to generate periodic train of a few picosecond wide pulses at high repetition rates. 

We have also shown that fractionalization is very sensitive to the range of interaction which 
governs the dispersion of edge magnetoplasmon eigenmodes: long range interactions might prevent observing fractionalization. 
Nevertheless, the dependence of the excess noise signal in the charge
of the pulses and in the propagation length gives information on the interaction range. Since our experimental setup could also be used to perform
high frequency admittance measurements in the $\nu=2$ edge channel system\cite{Bocquillon:2012-2}, 
it opens the way to ``on chip'' consistency checks of the edge magnetoplasmon scattering approach never realized before.

These results  argue for the development of experiments aimed at studying interaction induced electron/hole pair generation
in the $\nu=2$ quantum Hall edge channel system. In particular, performing such experiments on samples with side-gated
 copropagating edge channels as in recent experiments by F.~Pierre {\it et al}\cite{LeSueur:2010-1} 
 as well as on samples without gating might be a good way to access various interaction ranges. 
 
 An important issue of great interest is the 
 influence of edge smoothness on the electronic transport properties of the edge channels.
Several authors\cite{Aleiner:1994-1,Aleiner:1995-1,Han:1997-2,Johnson:2003-1}
have predicted the appearance of several branches of neutral excitations at the edge of a 2DEG in the integer quantum
Hall regime as well as dissipation for these all the gapless edge modes. But their existence has not been directly demonstrated so far. 
Studying electron/hole pair production along propagation of voltage pulses and performing high frequency admittance measurements
could help clarifying this issue and understand the dynamics of the all gapless modes living at the edge of the 2DEG in the
integer quantum Hall regime.
 
\acknowledgements{We would like to thank E. Bocquillon and G. F\`eve as well as F.~Pierre, P. Roche and F. Portier for stimulating 
and useful discussions. D. Ferraro and E. Thibierge are thanked for a careful reading of the manuscript. 
This work is supported by the ANR grant ''1shot'' (ANR-2010-BLANC-0412) and the ERC Grant "MeQuaNo".
}

\appendix
\section{Photo assisted transition amplitudes}
\label{appendix:analytics}

Here, we compute the photo-assisted transition amplitudes in presence of the periodic AC voltage
$V_{\mathrm{ac}}(t)$ for the Lorentzian and square pulses.

\subsection{Rectangular pulses}
\label{appendix:rectangular}

Let us consider square pulses of width $\tau_0\leq T$, defined for $|t|\leq T/2$ by $V(t)=-\alpha h/e\tau_0$ for $|t|\leq \tau_0/2$,
$V(t)=0$ for $|t|>\tau_0/2$. For a $T$-periodic train of such pulses, the Floquet amplitudes $C_l(\alpha,f\tau_{0})$ are given by:
\begin{eqnarray}
C_l(\alpha,f\tau_{0}) & = & (-1)^l\frac{\sin{(\pi(l+\alpha)(1-f\tau_0))}}{\pi(l+\alpha)}\nonumber\\
& + & f\tau_0\frac{\sin{(\pi\alpha-\pi(l+\alpha)f\tau_0)}}{\pi\alpha-\pi(l+\alpha)f\tau_0}\,.
\label{eq:square-pulse:1}
\end{eqnarray}
In the limit $f\tau_0\rightarrow 1$, these amplitudes vanish reflecting the fact that the ac part
of the voltage drive vanishes at $f\tau_0=1$. These amplitudes are also
obtained as $C_l=C(2\pi f(l+\alpha))$ where
\begin{equation}
\label{eq:square-pulse:2}
C(\omega)=\frac{2}{\omega T}\left(\pi\alpha\,
\mathrm{sinc}{\left(\frac{\omega\tau_0}{2}-\pi\alpha\right)}+\sin{\left(\frac{\omega T}{2}-\pi\alpha\right)}\right)
\end{equation}
where $\mathrm{sinc}(x)=\sin{(x)}/x$.
In the limit $f\tau_0\ll 1$, the main contribution to the partition noise arises from the squared modulus of
the first term in the r.h.s of \eqref{eq:square-pulse:2}. At zero temperature, the discrete sum \eqref{eq:HBT:excess-noise} is a regularized expression for the 
integral expression for the noise
\begin{equation}
\label{eq:square-pulse:3}
\Delta S_{1,1}^{(\mathrm{out})}\simeq \mathcal{RT}(\alpha e)^2f\int
\mathrm{sinc}^2{\left(\frac{\omega\tau_0-2\pi\alpha}{2}\right)}\,\frac{d\omega}{|\omega|}
\end{equation}
originally considered by Lee and Levitov\cite{Lee:1993-1}. As expected, this expression
exhibits an IR logarithmic divergence when $\alpha$ is not an integer which is a signature of the orthogonality catastrophe
associated with the fractional charge pulses\cite{Lee:1993-1}. For a periodic train of pulses\cite{Dubois:2012-1}, this divergence is regularized by
the period but manifests itself as a logarithmic divergence
of the outcoming excess noise $\Delta S_{1,1}^{(\mathrm{out})}$ in the limit $f\tau_0\rightarrow 0$. 
When $\alpha$ is integer, the logarithmic divergence at $\omega=0$ is replaced by a peak.

Note that the integrand in \eqref{eq:square-pulse:3} always has a peak for $\omega\tau_0=2\pi\alpha$ and
an $|\omega|^{-3}$ behavior at large $|\omega|$. 

\subsection{Lorentzian pulses}

 The amplitude $C_l(\alpha,q)$ to absorb $n$ photons can be computed as a contour integral
($q=e^{-2\pi f\tau_0}$):
\begin{equation}
C_l(\alpha,q)=\oint_{|z|=1} z^{l+\alpha-1}\left(\frac{1-qz}{z-q}\right)^\alpha\,\frac{dz}{2\pi i}\,.
\end{equation}
When $\alpha$ is a positive integer, the integrand has a pole of order $\alpha$ for $z=q$ and
when $l+\alpha\leq 0$ another pole of order $1-(l+\alpha)$ at $z=0$. Since for $l<-\alpha$ the
integral can be computed by closing the contour at infinity, this immediately shows
that $C_l(\alpha,q)=0$ for $l<-\alpha$. In the general situation where $\alpha>0$ is not
an integer, the integrand presents a branch cut singularity connecting $z=0$ to $z=q$ along the real
axis and another one connecting $z=1/q$ to $z=\infty$.

For $0<\alpha<1$, deforming the unit circle around the appropriate branch cut leads to expressions for the photo-assisted amplitudes
in terms of hypergeometric functions. Expanding these functions in series of $q^2$ and using the complement formula for the $\Gamma$
function leads to new series expansions of $C_l(\alpha,q)$ in $q$ in which the cancellation of $C_l$ for 
$l+\alpha<0$ and $\alpha$ integer is manifest. For $l+\alpha\geq 0$
\begin{equation}
\label{eq:amplitudes:above}
C_l(\alpha,q) = \alpha\sum_{k=0}^{+\infty} \frac{\Gamma(k+l+\alpha)\,q^{l+k}}{\Gamma(l+k+1)}\,\frac{(-q)^k}{k!\,\Gamma(1+\alpha-k)}
\end{equation}
whereas for $l+\alpha\leq 0$, 
\begin{equation}
C_l(\alpha,q) = \alpha\sum_{k=0}^{+\infty} \frac{\Gamma(l+k+\alpha)\,(-q)^{l+k}}{\Gamma(l+k+1)}\,\frac{q^k}{k!\,\Gamma(1+\alpha-k)}\,.
\label{eq:amplitudes:below}
\end{equation}
When $\alpha$ is a positive integer, due to the $\Gamma(1+\alpha-k)$  at the denominator, the sum over $k$ is truncated to $k\leq \alpha$. For the same reason, terms with $k\geq -1-l$ vanish when $l\leq -1$. Consequently $C_l(\alpha,q)$ is a polynomial in $q$ for $l+\alpha\geq 0$ and $C_l(\alpha,q)=0$ for $l+\alpha\leq 0$. Note that for non integer $\alpha >0$, the vanishing of terms with 
$k\geq \alpha +1$ does not happen: both expressions are full series in $q$ and therefore do not vanish.
In the case of a high compacity source $f\tau_0\gg 1$ or equivalently $q\rightarrow 0$ and consequently, 
all $C_l(\alpha,q)$ vanish for $l\neq 0$ and $C_0(\alpha,q)\rightarrow 1$ as expected since in this limit the
AC voltage vanishes.

\section{Electron and hole excitations}
\label{appendix:eh-count}

The operators counting the number of electrons and holes excitations with respect to the Fermi level are 
defined by:
\begin{eqnarray}
N_{e} & = & \int_{0}^{+\infty}c^\dagger(\omega)\,c(\omega)\,d\omega\\
N_{h} & = & \int_{-\infty}^0c(\omega)\,c^\dagger(\omega)\,d\omega\,.
\end{eqnarray}
Their averages can then be obtained in terms of the single electron and single hole coherences:
\begin{eqnarray}
\label{eq:number:electrons}
\langle N_{e}\rangle & = & \frac{iv_F}{2\pi}\int \frac{\mathcal{G}^{(e)}(t,t')}{t-t'+i0^+}\,dt\,dt'\\
\label{eq:number:holes}
\langle N_{h}\rangle & = & -\frac{iv_F}{2\pi}\int \frac{\mathcal{G}^{(h)}(t',t)}{t-t'-i0^+}\,dt\,dt'\,.
\end{eqnarray}
In an Hanbury Brown and Twiss experiment, the current noise in the outcoming arms
contains a contribution of the current noises within each of the two incoming channels  denoted here by $1$ and $2$ and a contribution $\mathcal{Q}$
due to two particle interferences\cite{Degio:2010-4}:
\begin{equation}
S_{1,1}^{(\mathrm{out})}=\mathcal{R}^2S_{1,1}^{(\mathrm{in})}+
\mathcal{T}^2S_{2,2}^{(\mathrm{in})}+\mathcal{RT}\mathcal{Q}
\end{equation}
where $\mathcal{R}$ and $\mathcal{T}$ respectively denote the energy independent reflexion and transmission probabilities
of the QPC used to partition the incident electron beams. The Hanbury Brown and Twiss contribution $\mathcal{Q}$
can be expressed in terms of the overlap of single electron coherences within the two incoming channels\cite{Degio:2010-4}:
\begin{equation}
\label{eq:HBT:overlap} 
\mathcal{Q}=(ev_F)^2\int (\mathcal{G}^{(e)}_{1}\mathcal{G}^{(h)}_{2}+
\mathcal{G}^{(h)}_{1}\mathcal{G}^{(e)}_{2})(t,t')\,dt\,dt'\,.
\end{equation}
Using the electron and hole coherences of the Fermi sea at zero temperature 
for $\mathcal{G}_{2}^{(e/h)}(t,t')$ shows that 
$\mathcal{Q}=e^2 \left(\langle N_{e}\rangle+\langle N_{h}\rangle\right)$, thus showing that low frequency noise
measurements lead to the measurement of the total number of excitations at zero temperature.
 
In the case of a voltage pulse, the single electron coherence is given by
\eqref{eq:coherence:driven} and in a similar way:
\begin{equation}
\label{eq:coherence:driven:holes}
\mathcal{G}_{1}^{(h)}(t',t)=\frac{i}{2\pi v_F}\,\frac{e^{i(\phi_{V}(t)-\phi_{V}(t'))}}{t-t'-i0^+}\,.
\end{equation}
where $\phi_{V}(t)=\frac{e}{\hbar}\int_{0}^{t}V(\tau)\,d\tau$.
Substituting \eqref{eq:coherence:driven} and \eqref{eq:coherence:driven:holes} into \eqref{eq:HBT:overlap} leads to the
following expression for the total number of excitations:
\begin{equation}
\langle N_{e}+N_{h}\rangle = \int \kappa(t-t')
e^{i(\phi_{V}(t)-\phi_{V}(t'))}\,dt\,dt'
\end{equation}
where the kernel $\kappa(t-t')$ is given by
\begin{equation}
\kappa(\tau) = \frac{1}{(2\pi)^2}\left(
\frac{1}{(\tau+i0^+)^2}+\frac{1}{(\tau-i0^+)^2}\right)\,.
\end{equation}
Minimizing this quantity is precisely the problem solved by Levitov, Lee and Lesovik in their 1996 paper\cite{Levitov:1996-1}.

\medskip

In the case of a $T$-periodic source, this discussion must be adapted by considering a time average over a single period. 
Since experiments are performed at non zero temperature, the HBT contribution to the noise signals is altered by the anti bunching
of electron and holes excitations emitted by the source and the thermal electron and hole excitations emitted from
the second incoming channel. Here we are interested by the excess contribution due to the AC part $V_{\mathrm{ac}}(t)$ of the drive voltage.
Denoting the chemical potential of the second incoming channel by $\mu_{2}$ and its electronic temperature by $T_{\mathrm{el}}$, 
the excess HBT contribution is given in terms of the excess single electron coherence generated by the AC drive:
\begin{equation}
\Delta\mathcal{Q}=e^2\int_{-\infty}^{+\infty} \tanh{\left(\frac{\hbar\omega-\mu_{2}}{2k_BT_{\mathrm{el}}}\right)}\,
\frac{v_F}{2\pi}\Delta \mathcal{G}_{1,0}^{(e)}(\omega)\,d\omega\,.
\end{equation}
When $\mu_2=\mu_1$ (here conveniently set to zero), equation \eqref{eq:HBT:noise-signal} can then
be expressed in terms of the excess spectral density of electron and hole excitations $\Delta \bar{N}_e(\omega)$
and $\Delta \bar{N}_h(\omega)$ emitted per period by the source at energy $\hbar\omega>0$:
\begin{equation}
\label{eq:HBT:complement}
\frac{\Delta\mathcal{Q}}{e^2f}=\int_0^{+\infty} \tanh{\left(\frac{\hbar\omega}{2k_BT_{\mathrm{el}}}\right)}\,
\Delta(\bar{N}_e+\bar{N}_h)(\omega)\,d\omega\,.
\end{equation}
since these excess electron and hole per period spectral densities are related to the single electron
coherence by\cite{Degio:2010-4}:
\begin{eqnarray}
\Delta\bar{N}_e(\omega) & = & \frac{v_F}{2\pi f}\,\Delta\mathcal{G}_{1,0}^{(e)}(\omega)\\
\Delta\bar{N}_h(\omega) & = & -\frac{v_F}{2\pi f}\,\Delta\mathcal{G}_{1,0}^{(e)}(-\omega)\,.
\end{eqnarray}

\section{Finite temperatures and interactions}
\label{appendix:finite-temperature}

At non zero temperature $T_{\mathrm{el}}$, an Ohmic contact emits
an equilibrium state which can equivalently be described both in terms of electrons and in terms
of edge magnetoplasmon excitations. In the latter description, this state is the thermal
state at temperature $T_{\mathrm{el}}$ of the bosonic excitations above the Fermi vacuum $|F_{\mu+\alpha hf}\rangle$ which is a vacuum state
for the edge magnetoplasmon modes. 

This thermal state can also be understood as a Gaussian statistical mixture of edge
magnetoplasmon coherent states $|[\delta\Lambda(\omega)]\rangle$. This Gaussian distribution
is centered around the vacuum $\Lambda(\omega)=0$ for all $\omega>0$. Its fluctuations  
$\overline{\delta\Lambda(\omega)\delta\Lambda^*(\omega')}$ are then equal to the thermal
average of the edge magnetoplasmon number $\langle b^\dagger(\omega)b(\omega')\rangle$ in
the thermal equilibrium state at temperature $T_{\mathrm{el}}$:
\begin{equation}
\label{eq:thermal:fluctuations}
\overline{\delta\Lambda(\omega)\delta\Lambda^*(\omega')}_{T_{\mathrm{el}}}
= \delta(\omega-\omega')\bar{n}(\omega,T_{\mathrm{el}})
\end{equation}
where $\bar{n}(\omega,T_{\mathrm{el}})$ denotes here the Bose-Einstein occupation number
$1/(e^{\hbar\omega/k_BT_{\mathrm{el}}}-1)$.
 
As in quantum optics, applying a time dependent drive voltage to the Ohmic contact corresponds to applying a displacement
operator to this thermal state. 
The resulting displaced thermal state can then be described
as a Gaussian statistical mixtures of edge magnetoplasmon coherent states around $\Lambda_V$ given by 
eq.~\eqref{eq:coherent:voltage-relation} and with fluctuations given by \eqref{eq:thermal:fluctuations}. 

To understand the action of the interaction region onto two such displaced
thermal states, let us first consider two incoming coherent states 
from the statistical ensembles defining the two displaced thermal states. They 
are characterized by their infinite dimensional complex parameters
$\Lambda_a^{(\mathrm{in})}$ with $a=1$ or $2$. The outcoming states is then a tensor product
of two coherent states characterized by 
\begin{equation}
\left(
\begin{array}{c}
\Lambda_1^{(\mathrm{out})} \\
\Lambda_2^{(\mathrm{out})} 
\end{array}\right)
= S\ldotp \left(
\begin{array}{c}
\Lambda_1^{(\mathrm{in})} \\
\Lambda_2^{(\mathrm{in})} 
\end{array}\right)
\end{equation}
Decomposing the incoming coherent state parameters into their averages and their Gaussian fluctuations
$\Lambda_{a}^{(\mathrm{in})}=\Lambda_a+\delta\Lambda_a$ ($a=1,2$) shows that, for each outcoming chanel,
we have a Gaussian statistical mixture of coherent states. The corresponding Gaussian distribution of the complex parameters
is centered around the classical outcoming parameters obtained by applying the edge magnetoplasmon scattering
matrix onto the vector of incoming parameters $(\Lambda_a)_{a=1,2}$:  this is nothing but the scattering
for wave amplitudes by a frequency dependent beamsplitter in optics. 
The outcoming fluctuations in each channel are Gaussian but nevertheless do not correspond to any thermal
equilibrium as noticed by Kovrizhin and Chalker\cite{Kovrizhin:2010-2}. The outcoming fluctuations
are given by:
\begin{eqnarray}
\langle b_a^\dagger(\omega)b_{a'}(\omega')\rangle_c^{(\mathrm{out})} & = & \delta_{a,a'} \delta (\omega-\omega')\,
\bar{n}_{\mathrm{out}}(\omega)\\
\bar{n}_{\mathrm{out}}(\omega) & = &
\sum_{b} |S_{a,b}(\omega)|^2\,\bar{n}(\omega,T_b)
\end{eqnarray}
where the $c$ subscript stands for "connected" and $T_b$ denotes the temperature of the incoming channel $b$.
When all the incoming channels have the same temperature $T_{\mathrm{el}}$, a thermal distribution 
is recovered due to the unitarity of the edge magnetoplasmon scattering matrix: $\sum_b|S_{a,b}(\omega)|^2=1$.

\section{Magnetoplasmon scattering for long range interactions}
\label{appendix:discrete-elements}

Let us denote by $Q_{j}(t)$ the charge stored within the interacting region of edge channel $j$. 
These charges are determined from the electrostatic potential $U_{j}(t)$ seen by electrons propagating in edge channel $j$ within the interacting
region through a capacitance matrix:
\begin{equation}
\label{eq:Butt:Q}
\left(
\begin{array}{c}
Q_{1}(t)\\
Q_{2}(t)
\end{array}\right)=
\left(
\begin{array}{cc}
C_{1} & -C \\
-C & C_{2}
\end{array}
\right)\ldotp\left(
\begin{array}{c}
U_{1}(t)\\
U_{2}(t)
\end{array}\right)\,. 
\end{equation}
Assuming both edge channels have the same Fermi velocity $v_{F}$, 
the dynamics of channel $j$ is determined by the equation of motion
\begin{equation}
\label{eq:Butt:eqs-motion}
(\partial_{t}+v_{F}\partial_{x})\phi_{j}(x,t)=\frac{e\sqrt{\pi}}{h}\,f(x)U_{j}(t)
\end{equation}
where $f(x)=1$ for $0\leq x\leq l$ and zero otherwise. The strategy is then to go to the frequency domain and 
rewrite eqs. \eqref{eq:Butt:Q} and \eqref{eq:Butt:eqs-motion} 
in terms of the incoming and outcoming bosonic field amplitudes and to eliminate the potentials $U_{j}(\omega)$. 
More precisely, in order to rewrite \eqref{eq:Butt:Q} in terms of the incoming and out coming bosonic fields,
one has to use the relation between the electronic density within each edge channel and the 
corresponding bosonic field: $:\psi^\dagger_j\psi_j(x):=(\partial_x\phi_j)(x)/\sqrt{\pi}$. Because \eqref{eq:Butt:Q}
involves the capacitance matrix, eliminating the  potentials $U_{j}(\omega)$ is most
conveniently performed after diagonalizing $\mathcal{C}$ with a rotation
$R_\theta=\cos{(\theta/2)}\mathbf{1}-i\sin{(\theta/2)}\sigma^y$. This finally leads to the scattering matrix:
\begin{equation}
\label{eq:two-channels:scattering:long-range}
S(\omega,l)=R_{\theta}^{-1}\,
\left(
\begin{array}{cc}
\mathcal{T}_{+}(\omega,l) & 0 \\
0 & \mathcal{T}_{-}(\omega,l)
\end{array}\right)
R_{\theta}
\end{equation}
where the angle $\theta$ is given by
($\Delta C=C_1-C_{2}$):
\begin{equation}
\label{eq:two-channels:cos-sin}
\exp{(i\theta)}  =  \frac{\Delta C/2+i\,C}{\sqrt{ C^2 + \frac{\Delta C^2}{4}}}\,
\end{equation}
and the phases $\mathcal{T}_{\pm}(\omega,l)$ have a non linear (dispersive) $\omega$ dependence given by:
\begin{equation}
\label{eq:eigen-transmissions}
\mathcal{T}_{\pm}(\omega,l)=e^{i\omega l/v_F}\,\frac{i\omega R_{K}C_{\pm} +1-e^{-i\omega l/v_{F}}}
{i\omega R_{K}C_{\pm} -1 + e^{i\omega l/v_{F}}}\,
\end{equation}
where the eigenvalues of the capacitance matrix are 
\begin{equation}
C_{\pm}=\frac{C_{1}+C_{2}}{2}\pm \sqrt{C^2+\frac{\Delta C^2}{4}}\,.
\end{equation}
The exponential $e^{i\omega l/v_F}$ in the r.h.s. of \eqref{eq:eigen-transmissions} 
represents the effect of free propagation and the other part contains the effect of the capacitances. Note that the free propagation limit
is obtained for infinite capacitances $C_{\pm}$. At low frequencies,  the plasmon eigenmodes become dispersionless: 
$\mathcal{T}_{\pm}(\omega,l)\simeq e^{i\omega l/v_\pm}$ where
the velocities $v_{\pm}$ are given by
\begin{equation}
\label{eq:long-range:velocities}
v_\pm  =  v_F+\frac{l}{R_{K}C_{\pm}}\,.
\end{equation}
Capacitances being proportional to $l$, the velocities $v_{\pm}$ are
independent from $l$: $v_{\pm}=v_F+(R_{K}\partial_{l}C_{\pm})^{-1}$. Therefore, in the infrared 
regime, the plasmon scattering matrix is of the form \eqref{eq:two-channels:universal-scattering}. 
An explicit check shows that the admittance matrix satisfies charge conservation and gauge invariance if and only if the capacitance
matrix satisfies the same condition. In such a case, each edge channel is totally screened
by the other one within the interaction region.

\medskip

It is interesting to evaluate the order of magnitudes of the various capacitances. We consider each edge channel as 
a quasi 1D conductor of length $l$ very large compared to its transverse dimensions $w$ as well as to the distance $d$ between
the two conductors. Then, the capacitances are of the form
\begin{equation}
C_{i,j}=\frac{\varepsilon_r\varepsilon_0}l\,F_{i,j}(w,d)
\end{equation}
where $\varepsilon_{r}$ denotes the relative permittivity of the material and $F_{i,j}(w,d)$ is a dimensionless function
of $w$ and $d$ that depends on the geometry. The dimensionless ratios of the $RC$ times to the time of flight $l/v_F$
are then given by:
\begin{equation}
\label{eq:estimates:alphas}
\frac{v_FR_KC_{i,j}}{l}=\frac{\varepsilon_r}{4\pi\alpha_{\mathrm{qed}}}\,\frac{v_F}{c}\,F_{i,j}(w,d)
\end{equation}
where $\alpha_{\mathrm{qed}}$ denotes the fine structure constant. Using $v_{F}\simeq 3\times 10^5\ \mathrm{ms}^{-1}$ and
$\varepsilon_{r}\simeq 10$ for $\mathrm{AsGa}$, one gets a numerical prefactor $v_{F}\varepsilon_r/4\pi c\alpha_{\mathrm{qed}}\simeq 0.4$.

\medskip

Let us now consider the case where $C_{1}=C_{2}$ so that $\theta=\pi/2$. The model then only depends on the
diagonal capacitances $C_{d}$ and the mutual capacitance between the two channels $C$. If we furthermore assume
that perfect screening is closed to be achieved, $C_{d}$ and $C$ are close together. Introducing $C_{\pm}=C_{d}\pm C$, 
we see that $C_{-}\ll C_{+}$ and therefore the two
velocities \eqref{eq:long-range:velocities} satisfy $v_{+}\ll v_{-}$. Note that here $v_{-}$ is the velocity of the symmetric (charge) mode whereas
$v_{+}$ is the velocity of the antisymmetric (dipolar) mode which, as in the case of short range interactions, is the slowest one.
For example using \eqref{eq:estimates:alphas} and $F_{11}=F_{22}=1$, we find $v_{F}R_{K}C_{+}/l\simeq 0.8$
leading to $v_+/v_{F}\simeq 1.25$. Consequently, the slowest velocity is of the order of $v_{+}\simeq 4\times 10^5~\mathrm{ms}^{-1}$. 
Assuming $C_{d}-C\simeq 0.05\times (C_{d}+C)$, we have $v_{-}/v_{F}\simeq 26$
thus leading to $v_{-}\simeq 7.8\times 10^6~\mathrm{ms}^{-1}$, a value of the same order of magnitude than the experimentally measured edge magnetoplasmon velocities 
in ungated  samples at filling fraction $\nu=2$\cite{Kumada:2011-1}.

The numerical computations in section \ref{sec:HBT:numerics} are performed using the following
expression for the scattering phase in terms of the dimensionless parameter $x=\omega l/v_F$:
\begin{equation}
\mathcal{T}_{\pm}(\omega,l)=\frac{e^{ix}-1+i\alpha_\pm x\,e^{ix}}{e^{ix}-1+i\alpha_\pm x}\,
\end{equation}
where $\alpha_{\pm}=R_{K}C_{\pm}v_F/l=v_FR_{K}\partial_{l}C_\pm$. Note that
this expression is clearly not periodic in $x\rightarrow x+1$ whereas it is equal to one
whenever $x$ is an integer. Under the hypothesis $C_{1}=C_{2}$ and very good screening, $\alpha_{-}\ll \alpha_{+}$. Using the
above numerical estimations: $\alpha_{+}\simeq 0.8$ and $\alpha_{-}\simeq 0.04$.


\end{document}